\documentclass[conference,10pt]{IEEEtran}

\usepackage{booktabs}
\usepackage[table,xcdraw]{xcolor}
\usepackage[normalem]{ulem}
\useunder{\uline}{\ul}{}

\usepackage{graphicx}
\usepackage[cmex10]{amsmath}
\usepackage[]{subfig}
\usepackage[font=small]{caption}
\usepackage{float}
\usepackage[T1]{fontenc}
\usepackage{array}
\usepackage{hyperref}
\newcolumntype{L}[1]{>{\centering\arraybackslash}m{#1}}
\usepackage{etoolbox}
\usepackage[ruled,linesnumbered]{algorithm2e}
\SetAlCapNameFnt{}
\SetAlCapFnt{}
\SetAlFnt{\footnotesize}
\usepackage{xspace}
\usepackage{makecell}
\usepackage{lipsum}
\usepackage{cleveref}[2012/02/15]
\crefformat{footnote}{#2\footnotemark[#1]#3}

\newcommand{\action}{return path steering\xspace}

\makeatletter
\patchcmd{\@maketitle}
  {\addvspace{0.5\baselineskip}\egroup}
  {\addvspace{-0.6\baselineskip}\egroup}
  {}
  {}
\makeatother

\begin{document}

\title{{\huge Withdrawing the BGP Re-Routing Curtain:}\\ \Large Understanding the Security Impact of BGP Poisoning via Real-World Measurements}


\author{\authorblockN{Jared M. Smith, Kyle Birkeland, Tyler McDaniel, and Max Schuchard}
\authorblockN{University of Tennessee, Knoxville}
\authorblockA{\{jms,kbirkela,bmcdan16,mschucha\}@utk.edu}}

\IEEEoverridecommandlockouts
\makeatletter\def\@IEEEpubidpullup{6.5\baselineskip}\makeatother
\IEEEpubid{\parbox{\columnwidth}{
    Network and Distributed Systems Security (NDSS) Symposium 2020\\
    23-26 February 2020, San Diego, CA, USA\\
    ISBN 1-891562-61-4\\
    https://dx.doi.org/10.14722/ndss.2020.23240\\
    www.ndss-symposium.org
}
\hspace{\columnsep}\makebox[\columnwidth]{}}

\maketitle

\begin{abstract}
The security of the Internet's routing infrastructure has underpinned much of the past two decades of distributed systems security research. However, the converse is increasingly true. Routing and path decisions are now important for the security properties of systems built \textit{on top} of the Internet. In particular, BGP poisoning leverages the de facto routing protocol between Autonomous Systems (ASes) to maneuver the return paths of upstream networks onto previously unusable, new paths. These new paths can be used to avoid congestion, censors, geo-political boundaries, or any feature of the topology which can be expressed at an AS-level. Given the increase in use of BGP poisoning as a security primitive for security systems, we set out to evaluate the feasibility of poisoning in practice, going beyond simulation.

To that end, using a multi-country and multi-router Internet-scale measurement infrastructure, we capture and analyze over 1,400 instances of BGP poisoning across thousands of ASes as a mechanism to maneuver return paths of traffic. We analyze in detail the performance of steering paths, the graph-theoretic aspects of available paths, and re-evaluate simulated systems with this data. We find that the real-world evidence does not completely support the findings from simulated systems published in the literature. We also analyze filtering of BGP poisoning across types of ASes and ISP working groups. We explore the connectivity concerns when poisoning by reproducing a decade old experiment to uncover the current state of an Internet triple the size. We build predictive models for understanding an ASes' vulnerability to poisoning. Finally, an exhaustive measurement of an upper bound on the maximum path length of the Internet is presented, detailing how recent and future security research should react to ASes leveraging poisoning with long paths. In total, our results and analysis attempt to expose the real-world impact of BGP poisoning on past and future security research.
\end{abstract}

\section{Introduction}{\label{intro}}

Responsible for ensuring packets find their home once they begin their journey, the Internet routing infrastructure serves a key role in ensuring the reachability, availability, and reliability of online services. Given the importance of its fundamental role, the security of the routing infrastructure as a set of protocols and routing process has underpinned much of the past two decades of distributed systems security research. However, the converse is becoming increasingly true. Routing and path decisions are now important for the security properties of systems built \textit{on top} of the Internet.

In particular, research has begun to harness the de facto routing protocol on the Internet, the Border Gateway Protocol (BGP)\cite{RFC4271} and the functionality it provides to implement new offensive, defensive, and analytical systems. Growing in use, a method known as BGP poisoning has been leveraged by censorship circumvention, DDoS defense, and topology discovery. Rooted in the BGP RFC, BGP poisoning can be used by routing-capable entities. These entities are known on the Internet as Autonomous Systems, or ASes~\cite{ASRFC}. Fundamentally, BGP poisoning is now being used to maneuver an ASes' return traffic around specific AS-to-AS links, new regions of the Internet topology previously not visible to certain ASes, and other regions of interest. Critically, BGP poisoning and the re-routing it provides is being employed for \textit{security purposes}.

For example, Smith et. al. presented Nyx, a DDoS defense system that leveraged the ability to manipulate inbound traffic paths with BGP for a security purpose: to route around attacked links on the Internet~\cite{RAC}. Nyx relies on altering paths on the Internet to circumvent links affected by DDoS, and is evaluated on the entire Internet via simulation. Prior to Nyx, Katz-Basset et. al. demonstrated the use of BGP poisoning in practice for single link-failures, as opposed to DDoS-inflicted failures, with LIFEGUARD~\cite{KatzBassett:2012gd}. In the domain of censorship circumvention, decoy routing (DR) has become a standard means to avoid censoring eyes~\cite{Wustrow:tc,Wustrow:ua,Karlin:2011td,Houmansadr:uf,Bocovich:2016vu}; though, Schuchard et. al. presented the \textit{Routing Around Decoys} (RAD) attack and follow-on E-Embargos~\cite{RAD,Schuchard:2016wd}, which utilized the routing infrastructure to circumvent decoy routers by re-routing both outbound \textit{and} return paths to completely avoid decoy routers.  In general, these security systems rely on the real-world feasibility of BGP poisoning to carry out defensive security goals.

Yet other systems rely on the \textbf{opposite} assumption to provide security guarantees or to attack poisoning-enabled defenses, that BGP poisoning is \textit{infeasible} in the real-world. In response to Schuchard et al.'s RAD attack on decoy routing systems, Houmansadr et. al presented the \textit{Waterfall of Liberty} system, which demonstrated that RAD could be prevented with downstream decoys~\cite{Houmansadr:2014vt}, relying on altering return paths via BGP poisoning to be infeasible, and Goldberg et. al. added additional security guarantees to Waterfall~\cite{Bocovich:2018wu}. In the world of DDoS, Tran \& Kang et al. presented an adaptive Crossfire/Link-Flooding Attack (LFA)~\cite{tran2019feasibility} that challenged Nyx, which we term \emph{Feasible Nyx}, yet their approach is only measured in simulation, supported by passive observations gathered by a single major network. Additionally, Tran's claims about how ASes filter poisoned paths are supported by \textit{passive}, not active, measurements. These particular security systems rely on the \textit{infeasibility} of BGP poisoning to carry out both offensive and defensive security goals.

We recognize that the lack of evidence exposing BGP poisoning's real-world feasibility has given rise to diverging claims in the research community. Worse, even the network operator community is being affected, as multiple NANOG mailing list threads have led to long, heated debates regarding the feasibility, positives, and negatives of BGP poisoning as a re-routing mechanism~\cite{nanog1,nanog2,nanog3}. While some published research claims re-routing based on poisoning is feasible, like Nyx, RAD, and LIFEGUARD, other research, such as Waterfall and Adaptive LFA attacks, \textbf{assumes the opposite} claim is true when presenting their experiments and analysis. While this problem is unaddressed, network operators can not reasonably depend on BGP poisoning for defensive purposes, or refute the feasibility of and leave out BGP poisoning when designing threat models, pushing critical networks to lose out on the benefits of any of the systems described above. While deployment of poisoning is still possible without a ground truth of Internet behavior, no prior study outlines with real-world active measurements 1) how feasible re-routing with BGP poisoning is in practice, 2) how networks and ASes treat poisoned AS paths propagated by a Nyx defender, RAD adversary, or network operator, and 3) clarifies the security implications of 1 and 2. This paper serves as that study.

\noindent \textbf{Understanding and Analyzing Re-Routing Feasibility}\\
We present the first self-contained collection of novel, real-world, Internet-scale measurements that validate or refute assumptions made in simulation or passively in recent security literature, such as RAD, Waterfall, Nyx, and Adaptive CrossFire/Link-Flooding Attacks. We provide insight into utilizing BGP poisoning as a topology and congestion discovery tool. We re-evaluate the Nyx DDoS defense platform, examine the graph-theoretic aspects of return paths available to ASes, and build predictive models for ASes wishing to understand their vulnerability to poisoning. We examine not only the filtering of BGP advertisements using poisoning, but also what ASes and what policies are deployed by such ASes that filter. To understand whether routing operator groups walk the walk when it comes to poison filtering, we measure their behavior against ASes not in a popular security-first ISP consortium.

These findings were uncovered via a 6-month long series of active measurements, beginning in January 2018 until July 2018. These measurements employed an array of control-plane and data-plane Internet infrastructure. This infrastructure included a collection of ASes from our own organization and the broader Internet via PEERING~\cite{schlinker14peering}, a real-world BGP testbed, nearly all responsive and one-per-AS 5k traceroute sources from RIPE Atlas~\cite{RIPEATLAS}, and live streams of BGP announcements from RouteViews~\cite{routeViews} and the RIPE Routing Information Service~\cite{riperis}. In practice, we found that the Internet's treatment of BGP poisoning lies on a \textit{spectrum} of behavior when evaluated across 1,400+ experiments, conducted with permission and guidance from 5 geographically and topologically diverse ASes on the Internet.

Specifically, we find that for 1,460 instances of BGP poisoning, over 77\% of the distinct instances could be successfully maneuvered onto new, previously unreachable AS-links at some point in the original path. An average of 8 new links were discovered per path, for a total of 3 completely new paths on average. In 20\% of cases, more than 5 completely new paths were discovered, with a maximum of 19 unique paths in one particular case and 23 total new links in another case. Beyond enumerating new paths, we found that BGP poisoning can be used to route around 80\% of ASes with less than 2,500 customers, considered small to medium-sized transit ASes. By further refining these measurements, we uncovered additional cases of poison filtering for highly connected ASes such as L3 and Cogent. When poisoning downstream ASes in systems such as Nyx, RAD, and others, connectivity can be lost to the poisoned ASes if a less specific IP block is not advertised to cover for the poisoned AS. In our work, we conducted an assessment of the ability of a poisoning AS without a less specific covering prefix to maintain reachability to the poisoned ASes, finding that 30\% of all ASes on the Internet can reach a /25 prefix advertised by an AS with only a /24, which is critical to effective BGP poisoning in an IPv4-dominated Internet. Next, we investigated default routing on the Internet and found that for 36\% of ASes with only 2 providers (that is, multi-homed in the simplest case), even when the primary provider is poisoned, the AS will continue to route through it. This finding hints that placing Waterfall resistors nearest the last-mile yields benefits even in the presence of routing-capable censors. We set out to uncover the raw amount of poisoning that a routing-capable AS can carry out, which has direct implications for the effectiveness of systems such as Nyx in practice. We find that ASes can propagate paths up to 251 in length that are accepted by 99\% of the Internet via customer cone inference. Critically, we also find that the Nyx system performs roughly 30\% worse in practice than in simulation, and that routing working groups do "walk the walk", and do not only "talk the talk". Perhaps intuitively, the larger the AS or ISP, the more filtering of poisons occurs, and the smaller, the less filtering. Throughout the rest of this paper, we dive deeper into these and additional analysis of BGP poisoning's feasibility on the Internet. We summarize our results, key takeaways, impacts on existing security systems, and security ramifications in Table~\ref{table:experiment-summary}.

\begin{table*}[!ht]
\setlength\extrarowheight{2pt}
\centering
\caption{Experiment summaries and their takeaways, impacted security systems, and ramifications for Internet routing security in general.}
\footnotesize{
\resizebox{0.98\textwidth}{!}{%
\begin{tabular}{L{2.5cm}L{5cm}L{5cm}L{2.5cm}L{4cm}}
\toprule
\textbf{Experiment Conducted}                                                                            & \textbf{High-Level Description}                                                                                                                      & \textbf{Key Takeaways}                                                                                                                        & \textbf{Existing Security Systems Impacted} & \textbf{Security Ramifications}                                                                                                                                                                                                                                                                                                                 \\ \midrule
Section~\ref{path-enumeration}: Steering Return Paths                                                       & Explores which ASes can effectively conduct poisoning, as well as the properties of alternative paths & 77\% of cases with successful path steering, Avg. 3 new unique traversed paths, minimal poisons needed, $< 1\%$ latency increase for alternate paths               & Waterfall of Liberty, RAD, Nyx, Feasible Nyx, LIFEGUARD         & Real-world evidence supports the claims of poisoning-enabled systems, with caveats for specific topological cases \\
\rowcolor[HTML]{EFEFEF} 
Section~\ref{nyx}: Re-Evaluation of Nyx                                                                     & Re-evaluates the Nyx DDoS defense system with active measurements directly compared to simulated results                                                                   & Nyx performs 30\% less effective in practice, the inferred topology of the Internet used in simulation often does not match the topology and policies in practice & Nyx, Feasible Nyx, LIFEGUARD                                    & Success of a system in simulation and/or passive measurement \emph{does not guarantee} success (or the same findings) in the real-world                                                                                                                                                                                                                \\
Section~\ref{filtering-by-degree}: Filtering of Poisoned Advertisements                                     & Investigates the ASes that filter BGP poisoned advertisements as well as their relative size and other metadata                                                            & 80\% of ASes with less than 2,500 customers can be poisoned to 99\% of the Internet                                                                               & Waterfall of Liberty, RAD, Nyx, Feasible Nyx, LIFEGUARD         & For specific parts of the Internet topology, poisoning does not work very effectively, allowing systems that would otherwise be hampered by poisoning to thrive in specific regions of the Internet                                                                         \\
\rowcolor[HTML]{EFEFEF} 
Section~\ref{long-paths},~\ref{long-path-filtering}: Filtering of Long Poisoned Paths & Establishes an upper bound for the maximum path length able to be advertised on the Internet with BGP poisoning                                                            & Max path length of up to 255 ASes propagated to 99\% of the Internet                                                                                              & RAD, Nyx, Feasible Nyx, LIFEGUARD                               & Security systems which use poisoning have a fixed budget of poisons in reality, specifically 245 when factoring in the length of a normal AS path                                                                                                                                                                                                                                                                                                                                               \\
Section~\ref{defaults}: Declining Presence of Default Routes                                                & Discovers the prevalence and distribution of default routes on the Internet                                                                                               & For 1,460 samples, 55\% of fringe or no-customer ASes had default routes, while $< 10\%$ of transit ASes had default routes                                        & Waterfall of Liberty, RAD, Nyx, Feasible Nyx, LIFEGUARD         & Default routes do impact poisoning-enabled systems negatively, but can be avoided in specific topological cases, especially when the system is not deployed at the edge of the Internet                                                                                                                           \\
\rowcolor[HTML]{EFEFEF} 
Section~\ref{connectivity}: Growth of /25 Reachability                                                      & Uncovers how many ASes must lose reachability to poisoned ASes when leveraging BGP poisoning                                                                              & 56\% of observed ASes will propagate /25 prefixes and 31\% of ASes respond to traceroutes for a /25                                                               & RAD, Nyx, Feasible Nyx, LIFEGUARD                               & Reachability of /25 prefixes limits some systems using poisoning, but for most cases, poisoning-enabled systems claims hold up in an Internet that has changed greatly since the earliest measurements of /25 reachability                                                                                                                                                                                                                                                                                                                                        \\ \bottomrule
\end{tabular}%
}%
}
\label{table:experiment-summary}
\vspace{-10pt}
\end{table*}

\noindent \textbf{Our Contributions}
\vspace{-5pt}
\begin{itemize}
    \item We conducted the largest measurement study on BGP poisoning to date, comprising 1,460 successful/1,888 total poisoning cases. We publish our dataset, source code, and data analysis from the final results of this paper.~\footnote{\url{https://github.com/VolSec/active-bgp-measurement}} See Section~\ref{path-enumeration}.
   	 \item We reproduce recent security papers done in-simulation or passively, but now with active BGP poisoning on the live Internet. See Sections~\ref{nyx} and ~\ref{long-path-filtering}.
   	\item We constructed statistical models that serve as a first-step towards utilizing BGP poisoning as an AS operator \textit{without} requiring active tests or convincing senior IT administrators. See Section~\ref{predicting-ml}.
    \item We assess the extent and impact of poisoned path filtering from several perspectives. For this analysis, see Section~\ref{filtering}.
    \item We reassessed the Internet's behavior with respect to default routes and /25 reachability 1 decade after the first exploration. See Sections~\ref{defaults} and ~\ref{connectivity} for these findings.
    \item We discuss insights and recommendations for the use (or threat model inclusion) of poisoning in security and measurement work going forward. We cover these insights within Sections~\ref{path-enumeration},~\ref{filtering},and~\ref{reachability}, with summaries in subsections at the end of these sections.
    \item We conclude the paper with discussion in Section~\ref{discussion} about the reproducibility and limitations of our experiments.
\end{itemize}
\section{Background and Motivation}{\label{background}}
\subsection{The Border Gateway Protocol (BGP)}

\begin{figure*}[!ht]
	\centering
	\subfloat[Critical links congested (Nyx) or decoy router placed (RAD)]{\label{fig:frrp1}\includegraphics[width=0.24\textwidth]{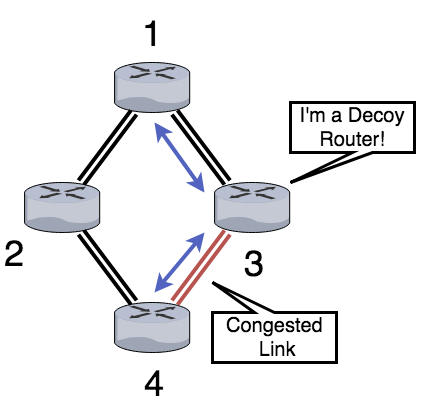}}
	\hspace{0.35cm}
	\subfloat[Lying about paths and prepending ASes to avoid]{\label{fig:frrp2}\includegraphics[width=0.27\textwidth]{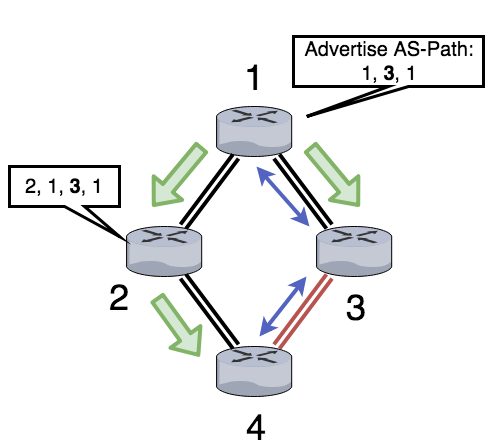}}
	\hspace{0.35cm}
	\subfloat[Loop detection triggered and new path taken]{\label{fig:frrp3}\includegraphics[width=0.33\textwidth]{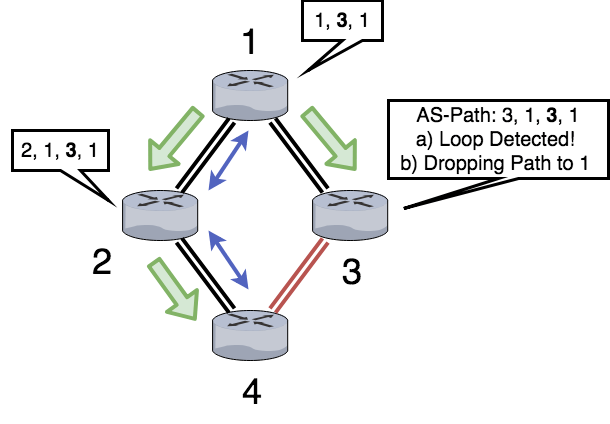}}
	\vspace{-10pt}
	\begin{center}
		\caption{Illustration of BGP Poisoning}{\label{fig:frrp}}
	\end{center}
	\vspace{-30pt}
\end{figure*}

The Internet is composed of many autonomous systems, or ASes, which are sets of routers and IP addresses under singular administrative control~\cite{ASRFC}. Each AS has one or more IP prefixes allocated to it, containing large amounts of IP addresses (e.g. an /8 or /16 subnet), or they can contain relatively few IPs (e.g. a /23 or /24). Today, a /24 is the most specific, or smallest, prefix recommended to be allowed by the most current best practices documents~\cite{rfc7454}. mThe Border Gateway Protocol~\cite{RFC4271} (BGP) is the de facto routing protocol of the Internet. BGP allows the exchange of information, called advertisements, between ASes about routes to blocks of IP addresses (e.g. prefixes), allowing each AS to have knowledge of how to forward packets toward their destinations. BGP advertisements are confined to the control-plane of the Internet, while protocols such as TCP and UDP are confined to the data-plane.

To carry out the routing decision process, BGP harnesses a path-vector routing algorithm \textit{with policies} to build and propagate AS paths, or routes, via BGP advertisements. Individual routers can define their own policies for which routes are considered "best" and then use the preferred routes to forward packets. In practice, these routes are often not the shortest, but rely on the specific policies defined in router configurations. These can include the cheapest route, the most favorable for congestion directly upstream, or any number of preferences a network operator sets for which upstream AS should be used. Outbound AS-level BGP paths are controlled by using the local routing policy to force a particular installed route as the first choice. BGP also includes a "loop detection" mechanism, where a BGP router receiving a new advertisement will first scan the entire path, and if it is already on the path, will drop (ignore) the advertisement and refuse to propagate the path to its neighbors.

To stabilize the control plane, mechanisms such as route-flap dampening~\cite{rfc2439,rfc7196} (RFD) and Minimum Route Advertisement Interval (MRAI) timers~\cite{RFC4271} limit the number of advertisements a single AS can propagate to amounts capable of being handled by connected ASes. These mechanisms can slow the process of BGP convergence, or the time taken for the Internet to settle on a set of stable routes to destinations based on BGP updates. However, as router processing power has increased, RFD becomes less widely used and is now disabled by default in Cisco routers~\cite{rfd-use}. Additionally, RIPE recommends setting RFD with a high BGP update suppression threshold~\cite{rfd-use}. MRAI timers also vary widely in configuration, with a default value of 30 seconds. We discuss in Section~\ref{infrastructure} how our experiments account for the presence of RFD and MRAI timers with appropriate wait times between BGP advertisements.

\subsection{BGP Poisoning}

While network operators can control the path their outbound traffic takes, they cannot directly control the path their inbound traffic takes. However, BGP poisoning is a traffic engineering technique that allows network operators to indirectly control inbound traffic. Poisoning \emph{is not} a standardized behavior in BGP, rather it is a side-effect of the loop detection mechanism mentioned earlier.

In detail, AS-level operators may use \textit{BGP poisoning} to prevent one or more ASes from installing, utilizing, and propagating a particular path~\cite{RAC,KatzBassett:2012gd,Anwar:2015tk}. The AS utilizing the technique (poisoning AS) determines a set of ASes they want inbound traffic to route around (poisoned ASes). In order to do this, the poisoning AS inserts \textit{the poisoned ASNs} into the AS\_PATH. According to the BGP specification~\cite{RFC4271}, the poisoned ASes \textit{should} drop the poisoning AS's path because of BGP's aforementioned loop detection mechanism. 

We illustrate BGP poisoning in Figure~\ref{fig:frrp}. In~\ref{fig:frrp1}, AS 1 wishes to move the best path of AS 4 to AS 1 off of the link over AS 3. Some security-related reasons for this could include avoiding congestion outside the control of the victim AS, attacking censorship circumvention systems, or routing around privacy-compromising regions of the Internet. In Figure~\ref{fig:frrp2}, AS 1 will now advertise a new BGP path, but now including the AS to avoid prepended at the end of the advertised path. This is the "poisoning" of the link over AS 3 by AS 1. This path will then be seen by AS 2 and AS 3. In Figure~\ref{fig:frrp3}, this advertisement will propagate past AS 2 to AS 4, but will be \textit{dropped} at AS 3 due to BGP's loop detection mechanism, since AS 3 sees its own AS number on the path. Now that AS 3 drops the path, it no longer has a route to AS 1 and will not advertise the path to AS 1 over itself. At this point, the new return path swaps to the path via AS 2, completing this poisoning instance.

BGP poisoning allows an operator to indirectly control the inbound AS-level path for their prefixes, though other less effective mechanisms for inbound path control exist. The Multi-Exit Discriminator (MED)~\cite{MED} attribute can influence a neighboring AS's tie-breaking process, but routers only employ MED after LOCAL\_PREF and AS\_PATH length. This property still leaves the decision to use the route in the hands of the neighboring AS's operators. Other techniques such as self-prepending, employing overlapping prefixes to trigger longest-prefix matches, and applying communities to routes may have an effect, but all rely on the remote AS's local policy.

\subsection{How Does BGP Poisoning Impact the Internet's Security?}{\label{impact}}


There are certain security systems that \emph{directly use BGP poisoning} to achieve their stated goals. In addition, other security systems rely on certain AS path properties to provide security guarantees. If an adversary could choose routes used by these security systems via BGP poisoning, then the claims of these systems would be undermined.

In the realm of censorship, BGP poisoning has been used by Schuchard et al.~\cite{RAD} with \textit{Routing Around Decoys} (RAD) to attack censorship circumvention systems, specifically those predicated on Decoy Routing (DR). Decoy routing is a recent technique in censorship circumvention where circumvention is implemented with help from volunteer Internet autonomous systems, called decoys. These decoys appear to route traffic to a decoy destination, but instead form a covert tunnel to the actual destination to evade the censor. In the RAD paper, only outbound BGP paths were altered to allow censors to route around decoys, but inbound paths could also be altered to avoid decoy routers. In response to this approach to routing around decoys, work by Houmansadr et al.~\cite{Houmansadr:2014vt,Nasr:2017fy} presented defenses against RAD, including the \textit{Waterfall of Liberty}. Waterfall places decoy routers on return paths under the assumption that RAD adversaries can not control these paths. Our study exposes the relative invalidity of this assumption.

Following from Waterfall, additional work was done by Goldberg et al. and others~\cite{Bocovich:2018wu,Minaei:2018wk} built on top of the return path decoy placement; thus, literature continues to emerge while operating under assumptions not entirely true in practice. Arguing that RAD placement was infeasible financially, Houmansadr et al.~\cite{Nasr:2016tp} showed the costs of RAD in practice, while Gosain et al.~\cite{Gosain:2017ty} places decoy routers to intercept the most traffic. Both approaches could be circumvented when BGP poisoning works successfully at certain topological positions.

In particular, Smith et al. uses BGP poisoning to provide DDoS resistance with Nyx~\cite{RAC} and  Katz-Bassett et al. uses poisoning for link failure avoidance with LIFEGUARD~\cite{KatzBassett:2012gd}. Nyx uses poisoning to alter the return paths of remote ASes to a poisoning AS, in an attempt to route the remote ASes' traffic around Link Flooding DDoS Attacks. LIFEGUARD uses poisoning to route around localized link failures between cloud hosts in AWS. Despite their success in simulation and limited sample sizes in practice, these systems assumptions need expansion and further validation at a wider scale to be used effectively for network defense. Tran et al.'s~\cite{tran2019feasibility} feasibility study of Nyx raises issues with poisoning needed to steer traffic, but fails to evaluate their assumptions via real-world active measurements. Instead, they rely on passive measurement and simulation. Our findings demonstrate how the real-world limitations of BGP poisoning affect these systems, specifically when BGP path steering via poisoning is used in a defensive context.

Since BGP poisoning is a non-standard technique, it is not widely known how feasible it is across the entire Internet, and further it is not known how its real-world feasibility differs from its de facto feasibility when simulated. This lack of understanding directly impacts existing security systems. As a result, we need to understand the real-world feasibility of BGP poisoning to shed light on the validity of these security systems' claims.

\subsection{Poisoning, RPKI, and BGPsec}{\label{rpki}}

The IETF has worked to add capabilities to cryptographically validate BGP advertisements in order to mitigate fabrication of AS paths, but adoption of these capabilities has been slow\cite{gilad2017we,sriram2019resilient}.  Given that BGP poisoning functions by fabricating portions of the AS level path, these defenses potentially present complications for operators using BGP poisoning.  

Route Origin Validation (ROV) introduced the Resource Public Key Infrastructure (RPKI). RPKI distributes trusted AS-level certificates along with Route Origin Authorizations (ROA)---signed attestations that an AS is permitted to advertise a prefix~\cite{rfc6480,rfc6482}. To perform ROV, an AS will take each advertisement, query the RPKI for any ROA that matches the advertisement's prefix and length, and ensure that the last AS in the path matches the AS in a ROA. In the case that no ROA exists for the prefix, no validation can be performed.  BGP poisoning can conform with ROA/ROV by appending a valid origin AS to the end of the path~\cite{rfc6811}. This allows poisoning-enabled systems to function in the face of ROV, although with a longer path. In Section~\ref{infrastructure} we highlight how advertisements used in our experiments conform to ROV.

BGPsec, first proposed as ``S-BGP'' in 2000 and standardized in 2017, adds the capability for full path validation by ensuring that each AS in the path has explicitly authorized the advertisement of the route to the subsequent AS in the path~\cite{rfc8205}.  BGPsec, if fully deployed, prevents poisoning, even with the previously-mentioned ROV bypass. However as of this writing, no commercial implementations of BGPsec exist~\cite{sriram2019resilient}, though partial deployments continue~\cite{lychev2013bgp}.  In partial deployments of BGPSec, routers will simply prefer routes that conform to BGPSec validation over routes which do not, rather than strictly dropping non-conforming routes.  As a result, systems which use poisoning as a primitive, such as Nyx~\cite{RAC}, may continue to operate so long as a strict full global deployment of BGPSec is not realized.

\subsection{Key Terminology}

We use the following terms in the rest of this paper:\\
\noindent \textbf{Steered AS}: The steered AS is a remote AS whose traffic is steered by the \textit{poisoning AS} onto new paths revealed via poisoning.\\
\noindent \textbf{Steered Path}: Steered AS traffic is moved onto a new \textit{steered path} by the poisoning AS' advertisements.\\
\noindent \textbf{Poisoning AS}: The poisoning AS exerts control over the \textit{steered AS} for security, measurement, performance, or other purposes.\\
\noindent \textbf{Poisoned AS}: Poisoned ASes are those being \textit{prepended} to advertisements by the \textit{poisoning AS} to steer paths.
\vspace{-3pt}

\section{Experiment Infrastructure}{\label{infrastructure}}

\begin{figure}[!ht]
\centering
\includegraphics[width=0.70\columnwidth]{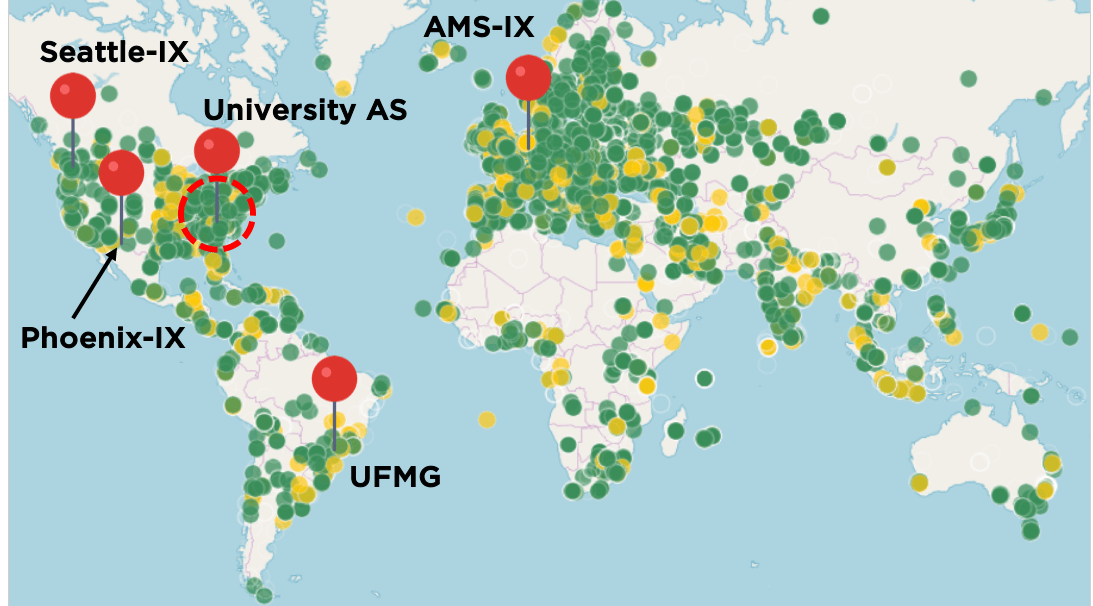}
    \caption{Distribution of RIPE Atlas traceroute probes at time of experiments with overlaid BGP routers}
	{\label{fig:atlas-peering}}
	{\vspace{-15pt}}
\end{figure}

The software-driven infrastructure used in our experiments to uncover the feasibility of BGP poisoning coordinates a vast amount of Internet infrastructure. We leverage thousands of network probes across 10\% of the ASes on the Internet and 92\% of the countries around the world, 5 geographically diverse BGP router locations \textemdash including two within Internet Exchange Points (IXPs) \textemdash and 37 BGP update collectors spread throughout the Internet. Our sample size of experiment vantage and measurement points represents the best available publicly at the time of the experiments. The major components of our measurement infrastructure are shown in Figure~\ref{fig:infra}. For a detailed discussion of our ethical considerations, please see the next section after we first cover our experiment infrastructure. We employ both existing and new network infrastructure in the \textit{control-plane} and \textit{data-plane}:

\begin{figure}[!ht]
    \centering
    \includegraphics[width=0.95\columnwidth]{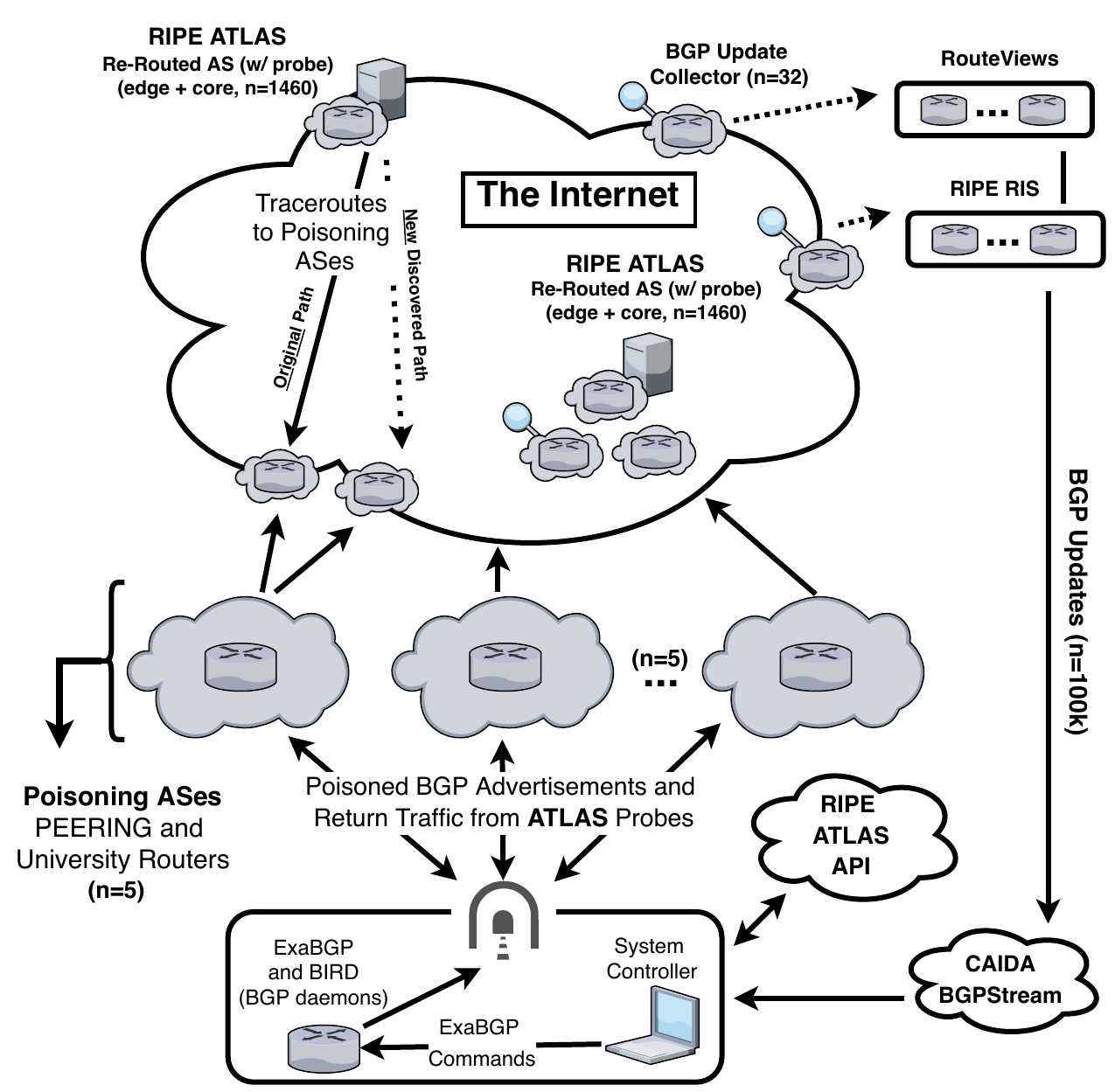}
    \caption{Measurement infrastructure from our experiments; incorporating CAIDA's BGPStream, RIPE Atlas, PEERING, a university AS, RouteViews, and RIPE RIS}
	{\label{fig:infra}}
	\vspace{-15pt}
\end{figure}

\noindent \textbf{Control-Plane Infrastructure}: We use BGP routers to advertise paths with poisoned announcements. The routers originate in a cooperating university AS, the University of Tennessee, Knoxville (AS 3450), and 4 routers from PEERING~\cite{schlinker14peering} advertised as AS 47065. Figure~\ref{fig:atlas-peering} shows the routers distributed both geographically and topologically across 3 countries: USA, Brazil, and the Netherlands. While this geographic diversity does not necessarily correspond with topological (i.e. AS-level) diversity), we used all available BGP routers from PEERING to generalize our results.

Advertisements were sent to 26 upstream transit ASes plus 300 peers. This includes two IXPs within PEERING. We employ 8 unused, unique /24 prefixes from PEERING and two /24 prefixes from the university AS. Active experiments pause 2 minutes between measurements after each BGP advertisement, and in some cases 10 minutes or more for different measurements depending on infrastructure constraints. These wait times help prevent route-flap dampening~\cite{rfc2439,rfc7196} and ensure expiration of MRAI timers~\cite{RFC4271}.

Poisoned advertisements are in the following format, which provides neighbor validation for the first AS and ROV for the last. This setup mirrors existing usage of AS-path prepending for traffic engineering use cases.

\begin{equation}\label{equ:frrp-adv}
	\resizebox{.4 \textwidth}{!}{
		$ \{\ AS_{origin},\ AS_{AV_{1}},\ AS_{AV_{2}},\ldots,\ AS_{AV_{N}}, \underbrace{AV_{origin}}_\text{For ROV} \}\ $
	}
\end{equation}

Finally, we monitor our BGP advertisements propagation from all 37 BGP update collectors available from CAIDA's BGPStream~\cite{Orsini:2016ug}. These collectors live physically within RouteViews~\cite{routeViews} and RIPE NCC's network~\cite{riperis}.

\noindent \textbf{Data-Plane Infrastructure}: We utilize RIPE Atlas~\cite{RIPEATLAS} to measure data-plane reactions to poison announcements. In total, we were able to conduct traceroutes across 10\% of the ASes on the Internet and 92\% of the countries around the world. We leverage RIPE Atlas's mapping of IPs to ASNs for discovering the AS-level path. For the path steering measurements using BGP poisoning, we only use 1 probe per AS, since we care about measuring new AS-level return paths, not router-level paths. We attempted to use every AS within the Atlas infrastructure as long as the probe was responsive and stable. We tried \textit{all probes available}, but only 10\% of ASes could be covered with responsive Atlas probes. While a system such as PlanetLab may also have been useful, PlanetLab has significantly smaller AS coverage compared to Atlas.

\noindent  {\bf Timing Considerations}: To ensure that our advertisements on the control-plane have propagated successfully by the time of traceroutes, our experiments wait at least 2 minutes between measurements after each BGP advertisement, and in some cases 10 minutes if conducted via PEERING infrastructure. These wait times help prevent route-flap dampening~\cite{rfc2439,rfc7196} and ensure expiration of minimum route advertisement timers~\cite{RFC4271}. We highlight additional data on BGP convergence times in the related work under Section~\ref{related-work}.
\section{Ethical Considerations}{\label{ethics}}

Our study conducts \emph{active} measurements of routing behavior on
the live Internet. As a result, we took several steps to ensure
that our experiments did not result in the disruption of Internet
traffic and were ethically sound. In particular, we ensure our experiments conform with the
Menlo Report~\cite{menloreport} and the policies of our infrastructure providers and
external network operators. To that end, we first engaged with the operator community and leveraged
their expertise throughout our experiments. Second, we designed
experiments to have minimal impact on routers and the normal network
traffic they carry. In this section we will touch on these steps.\\

\noindent \textbf{Working with Operators}: To ensure care was taken
throughout all experiments, we worked extensively with the network
operator responsible for campus-wide connectivity, quality-of-service,
and routing at the university AS used in our experiments. This
individual assisted in designing our experiments such that the
concerns of external network operators on the Internet would not be
affected adversely by our study, while also not biasing our
results. In addition to the university operator, we worked extensively
with PEERING operators from USC and Columbia throughout our study's
design and execution. PEERING operators have a large amount of
collective experience running active measurements on the Internet, 
which we leveraged to build non-disruptive
experiments.

Significant care was taken to \emph{notify} various groups of our
activities. In accordance with the PEERING ethics policy, we announced to
the RIPE and NANOG mailing lists prior to experiments the details of
the study, allowing operators the ability to opt out. Over the course
of our experiments we monitored our own emails and the mailing
lists. In total, 4 emails were received. Of the e-mails received, \textit{no parties asked to
  opt-out}. For each email received, we responded promptly, explained
our study, and incorporated any feedback provided.\\

\noindent \textbf{Minimizing Experimental Impact}: BGP path selection
is conducted on a per-prefix basis. Meaning that advertisements for a
particular prefix will only impact the routing of data bound for hosts
in that prefix. The prefixes used for our experimental BGP
advertisements were allocated either by PEERING or the university for
the \emph{express purpose of conducting these tests}. Outside of a
single host that received traceroutes, no other hosts resided inside
these IP prefixes. No traffic other than traceroutes executed as part
of our experiments were re-routed. This includes traffic for other IP
prefixes owned by the poisoning AS and any traffic to or from the
poisoned ASes.

Another potential concern is the amount of added, and potentially
unexpected, bandwidth load we place on links we steer routes onto.
Since the only traffic that was re-routed as a result of the
experiments was traceroute traffic bound for our own host, this added
traffic load was exceptionally low. The bandwidth consumed by our
measurements \emph{did not exceed 1 Kbps at peak}.

Besides minimizing non-experimental traffic, we minimized the impact
our BGP advertisements had on the routers themselves. Our BGP
advertisements were spaced in intervals also ranging from \textit{tens
  of minutes up to hours}. Resulting in a negligible increase to
router workloads given that on average BGP routers currently receive
16 updates per second during normal operation~\cite{potaroo}. All
updates were withdrawn at the conclusion of each experiment,
preventing unnecessary updates occupying space in routing
tables. Furthermore, all BGP updates conformed to the BGP RFC and were
not malformed in any way.

The largest concern to operators were our experiments measuring the
propagation of long paths on the Internet, described earlier in
Section~\ref{long-paths}. Several historical incidents, most notably
the SuproNet incident~\cite{supronet2009}, have demonstrated that
exceptionally long AS level paths can potentially cause instability in
BGP speakers. Underscoring this point were emails from operators on
the NANOG mailing
list~\footnote{\url{https://lists.gt.net/nanog/users/195871?search\_string=bill\%20herrin;\#195871}}
that appeared several months \emph{before} our experiments complaining
about instability in Quagga routers as a result of the advertisement
of AS paths in excess of 1,000 ASes. As a result, all experiments
involving long paths conformed to the filtering policies of our next
hop providers. In the case of PEERING experiments, administrators
limited our advertisements to 15 hops, and for the university, our
upstream providers (two large transit providers located primarily in
the United States) limited us to advertisements of 255 total ASes. These
limits were enforced with filters both in the experimental
infrastructure and at the upstream provider. In addition, such
experiments were conducted with 40 minute intervals between
announcements in an effort to allow operators to contact us
in the case of any instability resulting from our experiments.
\section{Feasibility of Steering Return Paths}{\label{path-enumeration}}


Our first set of experiments explore the degree to which a poisoning AS can, in practice rather than simulation, change the best path an arbitrary remote AS uses to reach the poisoning AS. We call this rerouting behavior \textit{\action}. Many security-related reasons for an AS to utilize \action focus on finding paths which avoid \emph{specific ASes}. As a consequence, we are interested in more than simply \emph{if} an AS can steer returns paths. We are interested in the diversity of paths available to a poisoning AS, the graph-theoretic characteristics of new usable paths, and to understand where such return paths play into security systems. In this section we both quantify the number of potential return paths we can steer a remote AS onto and dive deeply into the properties of these alternative paths. This analysis includes quantifying the diversity of transit ASes along those alternative paths, computing weighted and unweighted minimum cuts of the topology based on AS properties, and exploring latency differences between alternative paths. We also attempt to reproduce past security research and build statistical models that represent how successfully an AS can conduct \action.

\subsection{Experimental Design and Data Collection}

We explore the properties of alternative return paths by enumerating the paths a poisoning AS can move a remote AS onto via \action. Our set of poisoning ASes consisted of all ASes hosting a PEERING router plus the university AS. When conducting poisoning, the poisoning AS will only steer \textit{one remote/steered AS} at a time, where the remote AS is at least two AS-level hops away from the poisoning AS. This is critical to what we want to measure for security purposes. In this component of our study, we do not intend to measure new policies or congestion directly, as this has been done in prior work by Anwar et al.~\cite{Anwar:2015tk} which used multiple poisoning ASes from PEERING to steer the same set of remote ASes. However, Anwar et al.'s algorithm is fundamentally similar to ours. We use all available and responsive RIPE Atlas probes in unique ASes as steered AS targets. We collect BGP updates during the process in order to ensure our routes propagate and no disruption occurs. In total, we conducted our return path enumeration experiment for 1,888 individual remote ASes, or slightly more than 3\% of the IPv4 ASes that participate in BGP~\cite{potaroo}. We present the algorithm for the experiment below. The recursive function \textit{SteerPaths} builds a \textit{poison mapping}. This data structure maps the poisoned ASes required to reach a steered path. For all 1,888 instances, we capture the ASes and IPs of the original and all new paths, latencies at every hop, geographic IP locations, the set of poisoned ASes need to steer onto each successive path, and other relevant path metadata. This dataset will be made public upon acceptance. Our poisoning algorithm is measured to be successful when we see RIPE Atlas switch the path it is using to the poisoning AS. We infer that this sudden switch in path shortly after we confirm our poisoned BGP updates have propagated is due to the poisoning itself.

\renewcommand{\thealgocf}{}
{
\begin{algorithm}
    \SetKwInOut{Input}{Input}

    \underline{recursive function SteerPaths} $(src,dest,nextPoison,currentPoisons,mapping)$\:\\
    \Input{poisoning AS $src$, steered AS $dest$, next poisoned AS $nextPoison$, current poisons $currentPoisons$, poisonMapping $mapping$}
    $currentPath = src.pathTo(dest)$\\
    $poisonDepth = currentPath.indexOf(nextPoison)$\\
    $previousHop = currentPath[poisonDepth - 1]$\\
    $newPoisons =  currentPoisons + nextPoison$\\
    ~\\
    $dest.poison(newPoisons)$\\
    $currentPath = src.pathTo(dest)$\\
    $mapping.put(newPoisons, currentPath)$\\
    \If{$currentPath == \emptyset$}{
    $disconnected = true$\;
    }
    $newPrevHop = currentPath[poisonDepth - 1]$\\
    \If{$!disconnected\ \&\&\ newPrevHop == previousHop$}
{$SteerPaths(src,dest,currentPath[poisonDepth],$\\
$newPoisons,mapping)$\;
    }
    $dest.poison(currentPoisons)$\\
    $poisonIndex = currentPath.indexOf(nextPoison)$\\
    \If{$currentPath[poisonDepth\ + 1] == dest$}{$SteerPaths(src,dest,currentPath[poisonDepth + 1],currentPoisons,mapping)$\;
    }
    \caption{Recursive path steering algorithm}
    {\label{steering-algorithm}}
\end{algorithm}
}

\begin{figure*}[ht!]
    \centering
    \subfloat[Poisons Required to Discover Unique ASes and Entire Unique Return Paths with a Regression Line Fit]{\label{fig:avg-discovered-paths}\includegraphics[width=0.29\textwidth]{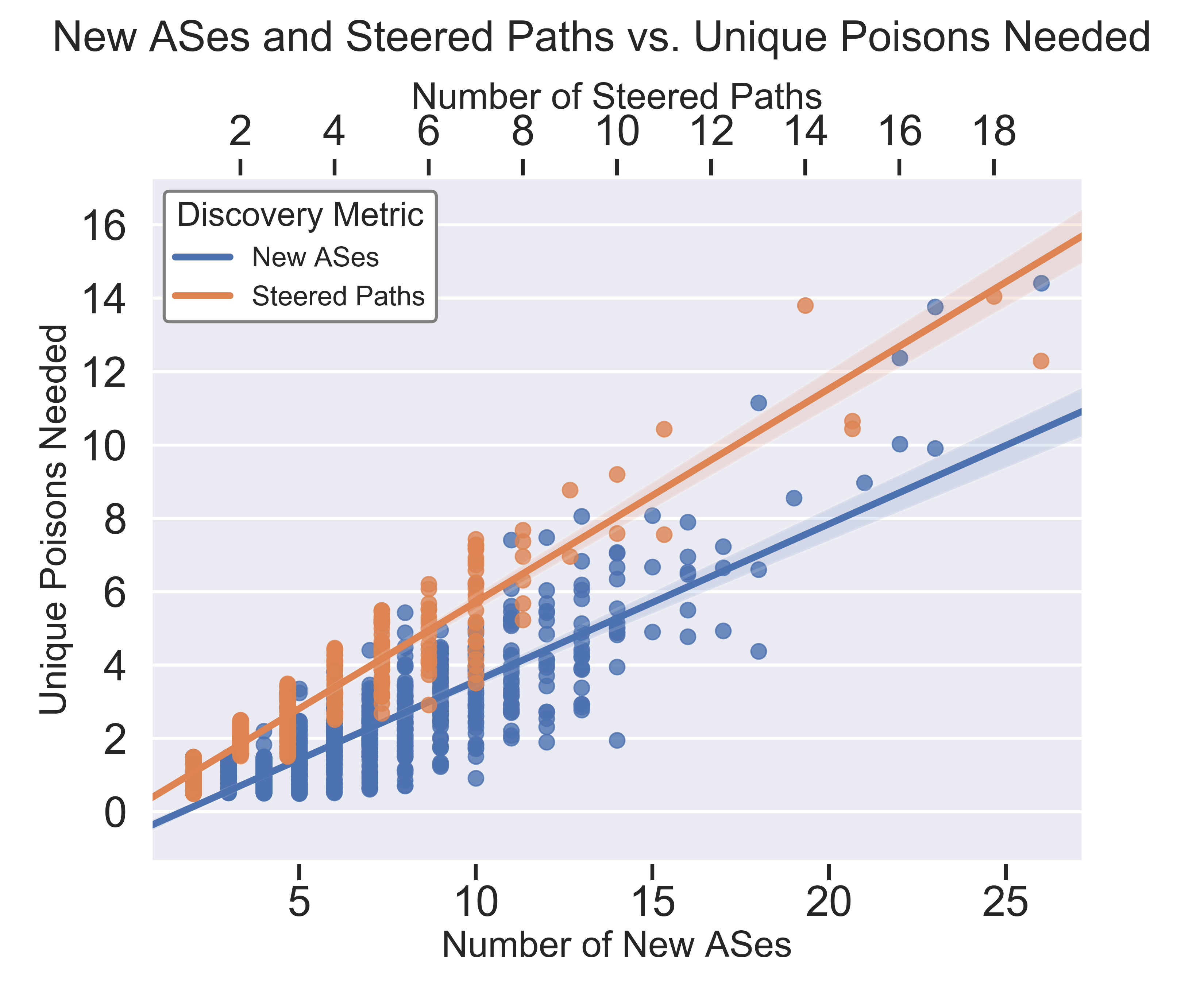}}
    \hspace{0.20cm}
    \subfloat[RTT Comparison, Original vs New Return Paths]{\label{fig:alternate-paths-rtt}\includegraphics[width=0.31\textwidth]{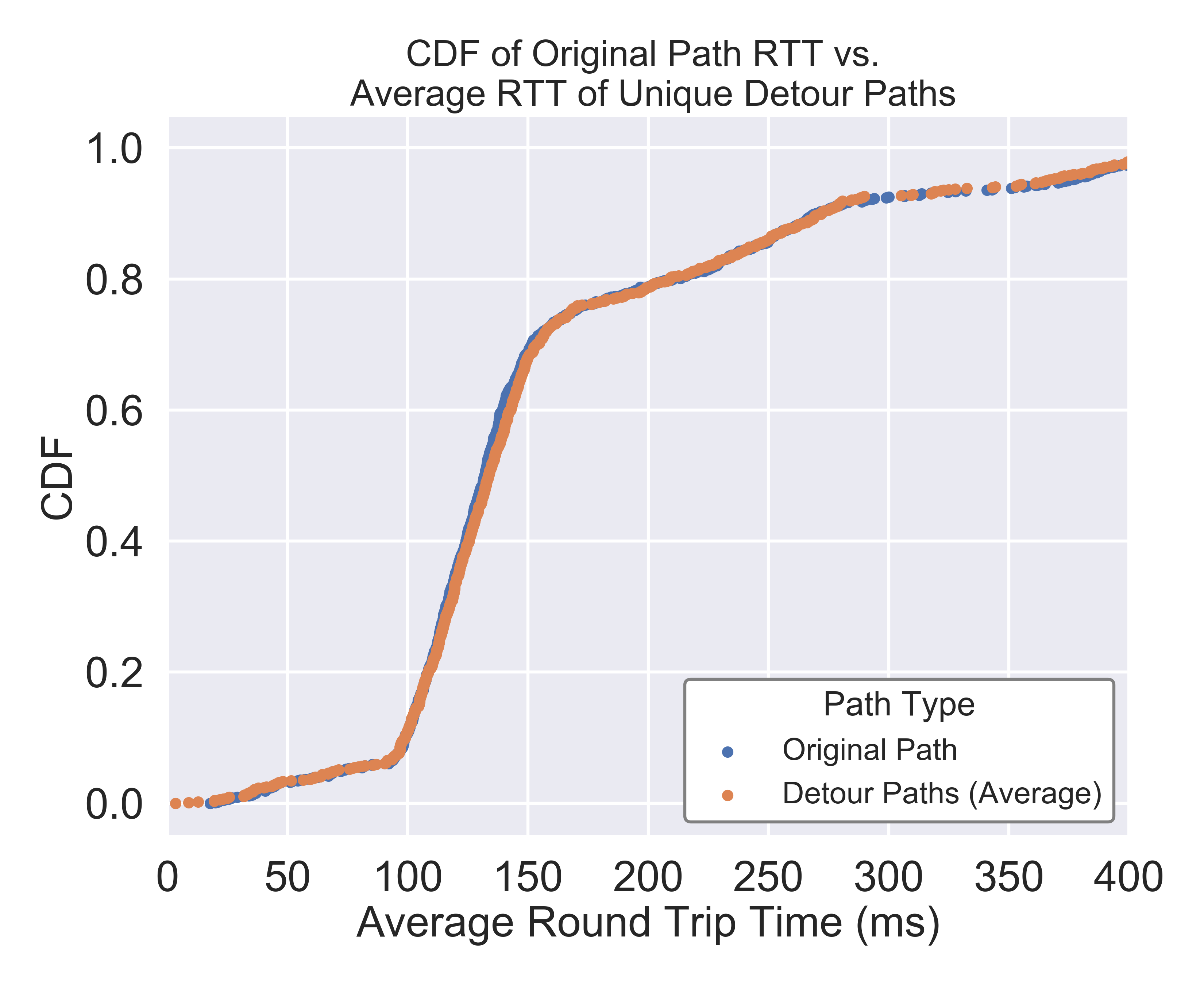}}
    \hspace{0.20cm}
    \subfloat[Active Measurement vs Simulation for an Identical Set of Poisoning-Steered AS Pairs]{\label{fig:rs-sim-vs-practice}\includegraphics[width=0.31\textwidth]{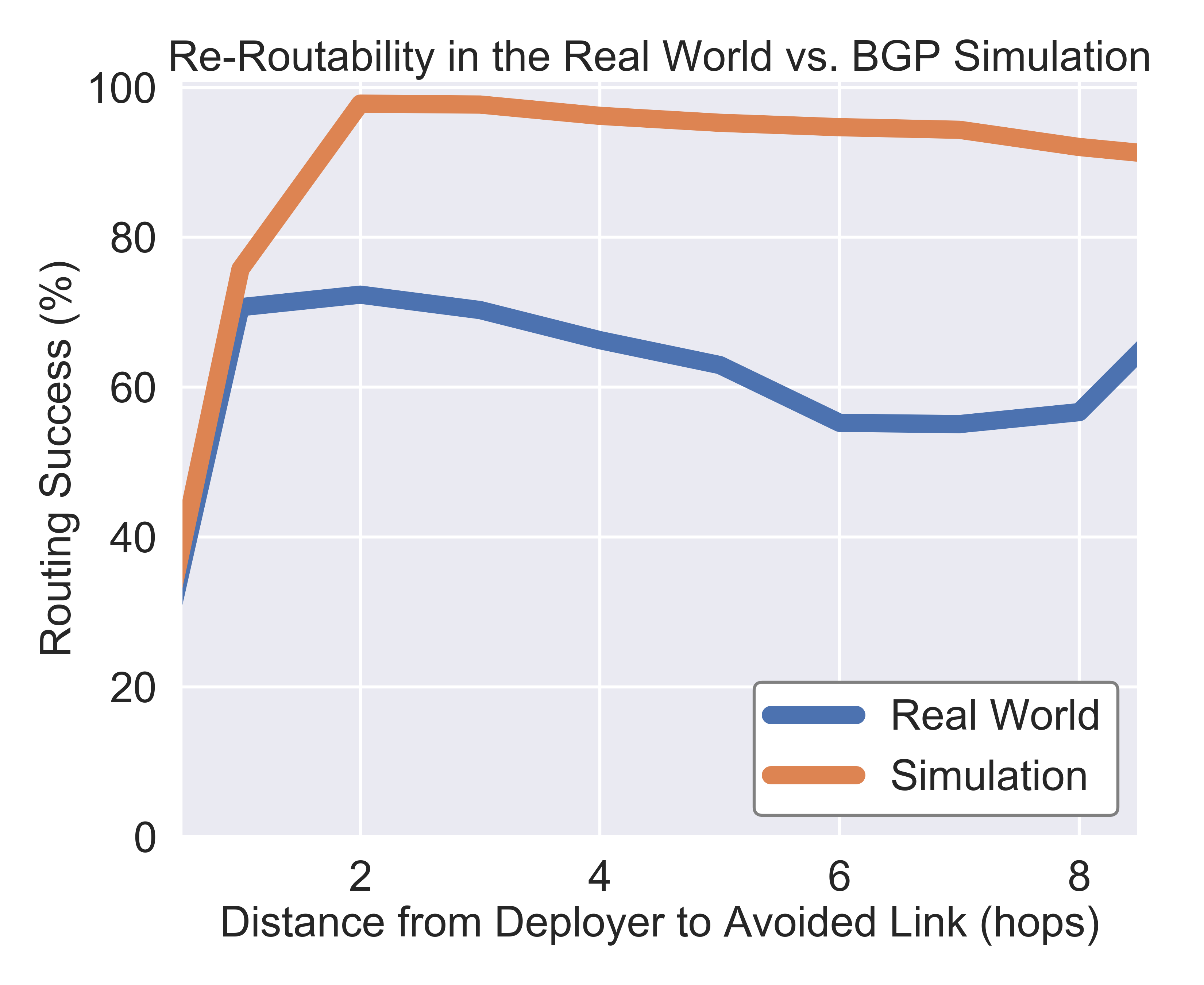}}
    \caption{Return path steering metrics. Figure~\ref{fig:avg-discovered-paths} shows the number of poisons required to reach steered paths. Figure~\ref{fig:alternate-paths-rtt} shows the difference in measured RTT between original paths and steered paths. Figure~\ref{fig:rs-sim-vs-practice} re-evaluates Smith et al.'s Routing Around Congestion defense}
    \label{fig:routing-around-ases}
    \vspace{-20pt}
\end{figure*}

\subsection{Steering Return Paths}

We successfully steered 1,460 out of 1,888 remote ASes, or 77\%, onto at least one alternative return path. The unsuccessful cases arose due to default routes (discussed in Section~\ref{connectivity}) or poison filtering (discussed in Section~\ref{filtering}). For each case of successful poisoning we analyzed several metrics: the number of unique alternate steered paths discovered and ASes traversed, the number of poisoned ASes needed to reach those paths, centrality measures of the graph formed by the steered paths, minimum cuts of this graph, and latency differences between the original path and the alternate return paths. Summary statistics of several of these measurements are shown in Table~\ref{table:path-metrics}.

\begin{table}[!ht]
\setlength\extrarowheight{2pt}
\centering
\resizebox{0.80\columnwidth}{!}{%
\begin{tabular}{ll}
\toprule
\textbf{Metric}        & \textbf{Result} \\ \midrule
Cases of Unsuccessful Return Path Steering & 428                                 \\
\rowcolor[HTML]{EFEFEF} 
Cases of Successful Return Path Steering   & 1,460                               \\
Overall Unique New ASes                    & 1369                                \\
\rowcolor[HTML]{EFEFEF} 
Average Unique Steered Paths Per Atlas AS  & 2.25                                \\
Average Unique New ASes Per Atlas AS       & 6.45                                \\
\rowcolor[HTML]{EFEFEF} 
Max Unique Steered Paths                   & 19                                  \\
Max Unique New ASes                        & 26                                  \\
\rowcolor[HTML]{EFEFEF} 
Avg. Poisons Needed vs. Avg. New ASes      & 2.03/6.45                           \\
Unique New ASes vs. Unique Poisons Needed  & 1369/468                            \\ \bottomrule
\end{tabular}%
}
\vspace{0.05cm}
    \caption{Summary of return path steering metrics}
\label{table:path-metrics}
\vspace{-15pt}
\end{table}

As shown by Figure~\ref{fig:avg-discovered-paths}, for three quarters of (steered, poisoning) AS pairs, between 2 and 3 unique paths were reached on average using \action. However, for some pairs, we find nearly 20 unique paths. Clearly, some ASes are better positioned to execute \action. We dive deeper into which ASes can more easily execute \action using an array of statistical and machine learning models later in Section~\ref{predicting-ml}. The number of poisons required to reach these paths scales linearly with both the number of discovered alternate paths and the number of unique new ASes on those paths. This is relevant for many systems relying on \action, as each poison increases the advertised path length by one. We will demonstrate in Section~\ref{filtering} that path length is a major factor in AS operators' decision to filter or propagate received advertisements.

A comparison of round trip times between original and steered paths as measured by traceroutes is shown in Figure~\ref{fig:alternate-paths-rtt}. The original and steered round trip time (RTT) values are nearly indistinguishable. We find that on average we only observed a 2.03\% increase in latency on alternative return paths, a positive indication that the alternative return paths have similar performance characteristics. Interestingly, we also found that the new paths tested out of the university AS performed 2.4\% \emph{better} than the original paths, while the steered paths out of PEERING ASes performed 4\% \emph{worse} than the original paths. We believe this is attributable to the proximity of the university AS to the Internet's core, versus the relative distance from the PEERING ASes to the core. These latency measurements provide supporting evidence that the alternative paths are fit to carry traffic from an approximate performance perspective, though the best indicator of path performance would come from knowing the bandwidth of the links traversed. Unfortunately, such data is highly sensitive and considered an industry trade secret for an ISP. Our approximation via the link round trip times is our best estimate to link viability, with more real-world applicability than the PeeringDB estimated bandwidth model approach used in simulation by work from Smith and Schuchard, as well as Tran and Kang~\cite{RAC, tran2019feasibility, Schuchard:2016wd}.

\subsubsection{Re-evaluation of Nyx}{\label{nyx}}

Next, we attempt to reproduce the performance of the Nyx Routing Around Congestion (RAC) system~\cite{RAC} with active measurements. Using the open source simulator from Smith et al.~\cite{chaos-sim}, we find 98\% routing success (ability to steer an AS around a congested link) for the same 1,460 samples we measured in our previous experiment.  We perform an exact comparison between simulated results and those measured actively. We find that in practice these ASes perform approximately 30\% \emph{worse}. We show these findings in Figure~\ref{fig:rs-sim-vs-practice} using the same metric from the Nyx paper for the simulation and in practice comparison.

This apples-to-apples comparison illustrates that in most cases \action functions in practice, but the extent of that functionality is not necessarily as substantial as simulations based purely on AS-relationship models~\cite{CAIDA} imply. In Nyx simulation, CAIDA AS-Relationship models were used to show that ASes have tens to hundreds of available paths based on paths observed via advertisements. This is \textit{significantly larger} than what we found in practice (~2-3 unique paths). This is due to the CAIDA AS-relationships dataset only attempting to show \textit{connectivity}, not policy. In other words, an AS-to-AS link observed in BGP advertisements does not indicate real-world willingness to send traffic over such links. While CAIDA's data represents the best possible model for simulation, it is clear that simulations relying on the routing infrastructure should be validated by active measurement. While Anwar et al.~\cite{Anwar:2015tk} find connectivity that CAIDA does not, we find that when considering only a single poisoning AS steering a single remote AS, the poisoning AS can not achieve the full spectrum of return paths shown by AS connectivity alone. We also find that a poisoning/steered AS pair in simulation often had a longer original path length than was measured in the real world. The simulator found no original paths where the length was 3 AS hops. For the same sample set actively measured, we found paths with an original path lengths of 3 hops in 165/1,460 successful steering cases.

Clearly, assumptions from the simulation of Nyx did not match what was discovered in reality. First, as stated before, inferred policies of Internet routes \emph{do not} match what is found in practice. Thus, both simulation of the inferred topology, and passive measurement of all paths seen, cannot directly justify individual policies that will actually be taken by a poisoning AS, which may contribute to the 30\% difference in success. Second, Nyx did not factor in the effects of ASes that filter poisons on the success of routing around congestion. While the simulation allowed paths to propagate without filtering, this is not true in practice for certain cases, as we discuss in the next Section. Third, Nyx did not limit the amount of new paths that could be used to re-route during simulation, while in practice this is restricted to ~2.5 available alternate paths. Therefore, the success from simulation will appear higher due to the lack of this restriction.

\begin{figure*}[ht!]
    \centering
    \subfloat[Average Normalized Steering Betweenness]{\label{fig:avg-diversity}\includegraphics[width=0.31\textwidth]{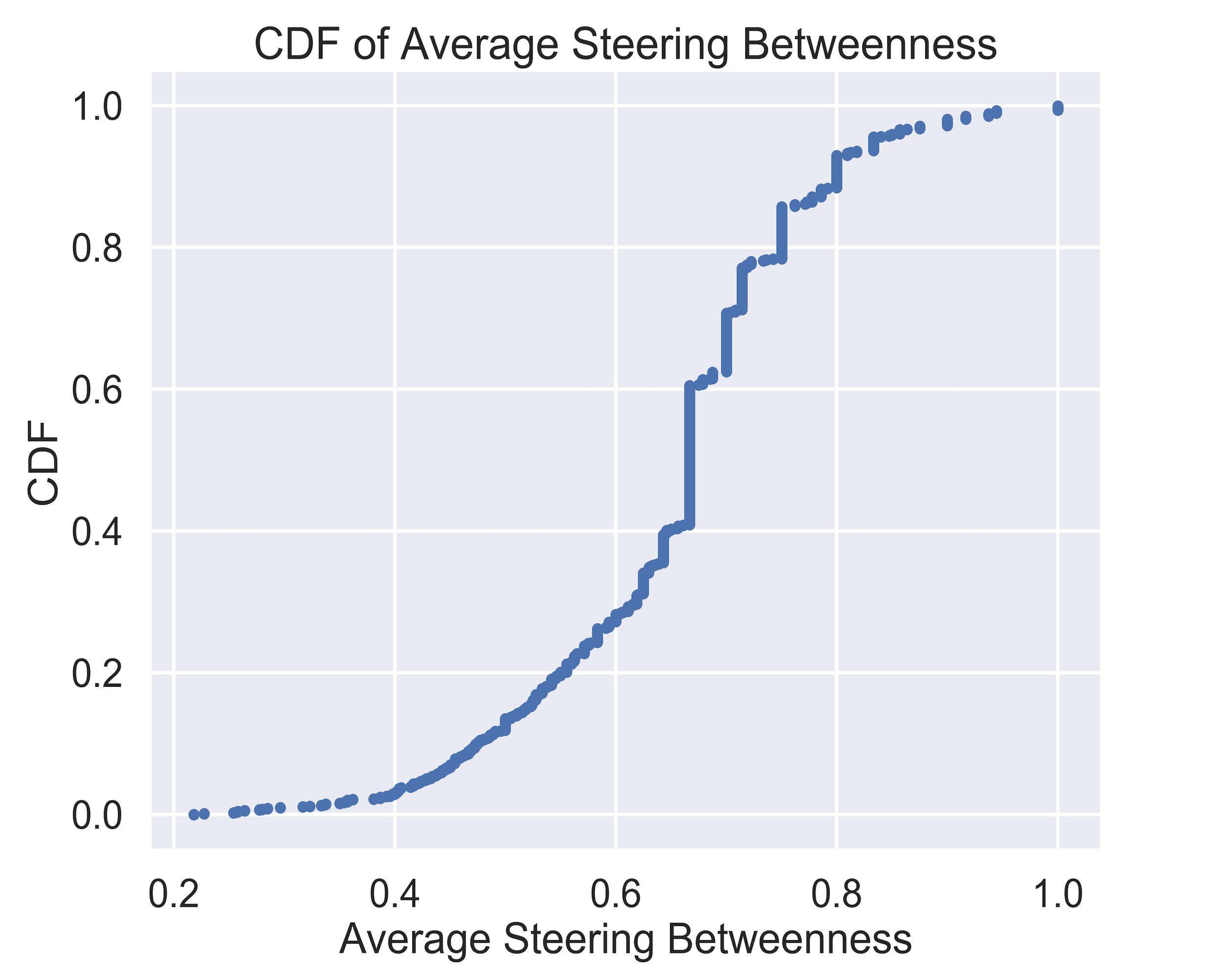}}
    \hspace{0.20cm}
    \subfloat[Unweighted Min Cut]{\label{fig:mincut-unweighted}\includegraphics[width=0.30\textwidth]{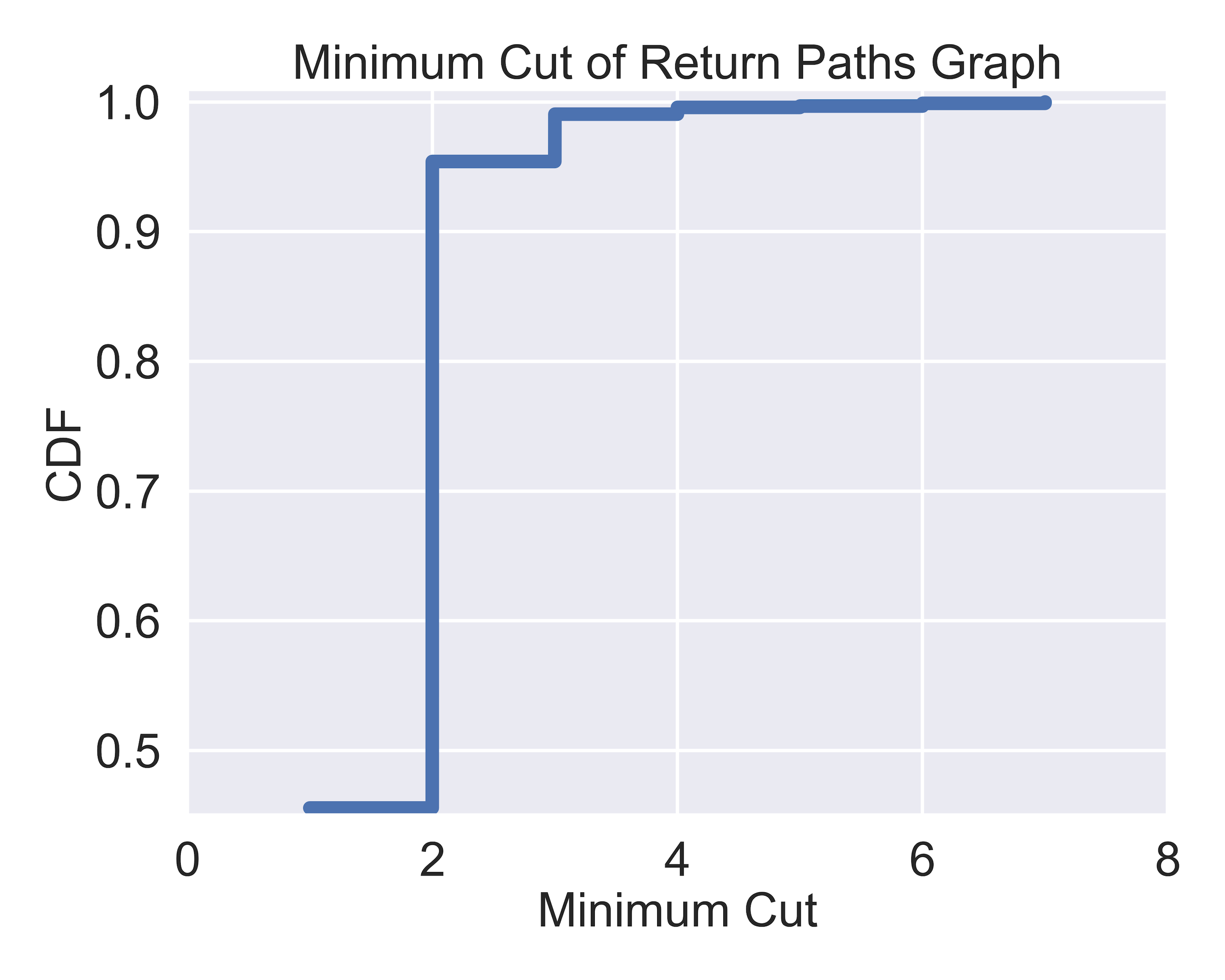}}
    \hspace{0.20cm}
    \subfloat[Weighted Min Cuts]{\label{fig:mincut-weighted}\includegraphics[width=0.30\textwidth]{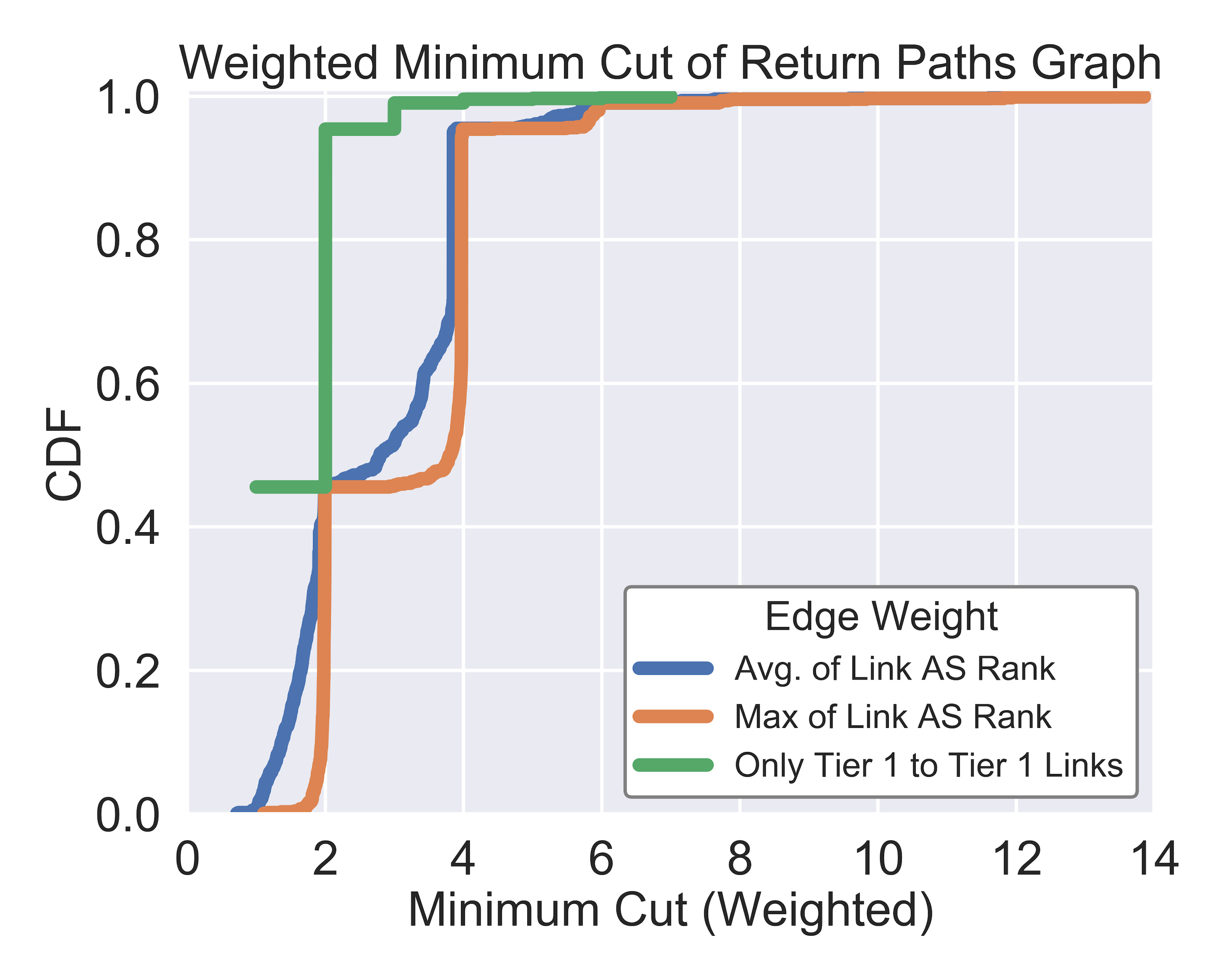}}
    \caption{Centrality measures of the importance of individual ASes in the directed acyclic graph formed by the original path and steered paths. Figure~\ref{fig:avg-diversity} shows the average vertex betweenness for ASes in each of the graphs, normalized by the number of distinct paths between steered and poisoning AS. Figure~\ref{fig:mincut-unweighted} and~\ref{fig:mincut-weighted} show the unweighted and weighted min cuts of these graphs}
    \label{fig:mincuts}
    \vspace{-20pt}
\end{figure*}

\subsubsection{Graph Theoretic Analysis of Return Path Diversity}{\label{graph-theory}}

Here we analyze characteristics of the directed aclycic graph formed by combining the original and alternate return paths from the steered AS to the poisoning AS across all steering experiments (1,888 instances). We call this the \textit{return path graph}.

One first concern is AS-level path diversity of the return path graph; how different are potential return paths are in terms of the AS-level hops they contain? This is relevant because security systems built on \action may seek to avoid specific links (e.g., to route around congestion). In this case, the availability of alternate return paths alone is not sufficient. The poisoning AS requires a return path \textit{that does not contain} the congested link. Here we quantify the diversity of return paths by calculating the average betweenness for AS nodes on the return path graph. For each AS in the graph, we count how many paths the AS appears on, and divide by the number of total return paths (original and all discovered alternates). This yields a normalized betweenness for each AS between 0 (exclusive) and 1. The average betweenness for ASes on the return path graph, which we call \textit{steering betweenness}, is designed to explore the diversity of ASes along the original and alternate return paths. A steering betweenness approaching 1 implies that the set of possible return paths differ in AS hops very little, while a number close to 0 implies that there are few ASes found in multiple return paths.

\noindent Figure~\ref{fig:avg-diversity} shows steering betweenness for each poisoning/steered
AS pair in our experiments. We see that on average, a transited AS from the return path graph has a betweenness centrality of roughly $0.667$. This indicates that some ASes appear on the majority of return paths. However, these paths are \textit{not} essentially identical.

Next, We also compute the unweighted and weighted minimum cut of the return path graph. Here we seek to explore the prevalence of bottlenecks, or links that can not be steered around, in the set of return paths. This metric is especially meaningful for systems like Nyx that use BGP poisoning to maintain connectivity between a selected AS (the steered AS in the context of this experiment) and a Nyx deployer (the poisoning AS) in the presence of a DDoS attack, since a low minimum cut reflects an unavoidable bottleneck for DDoS to target. Figure~\ref{fig:mincut-unweighted} demonstrates that in just under half of cases a single bottleneck exists, and for more than 90\% of steered/poisoning AS pairs, a bottleneck of at most two links exists in this graph.

To explore where in the topology these bottlenecks occur, we constructed different methods for weighting the graph, seen in Figure~\ref{fig:mincut-weighted}. First, we assign infinite weight to all Tier-1 to Tier-1 links to effectively remove them from consideration in the minimum cut, as the real-world capacity of links between large providers is, intuitively, much greater than links between other ASes. Consequently, we expect they are more difficult to degrade. Tier-1 ASes are those ASes who have no providers, and can therefore transit traffic to any other AS without incurring monetary costs~\cite{Oliveira:wn}. Interestingly, this did not change the minimum cut for any graph, meaning that the bottlenecks did \emph{not occur as a result of single unavoidable Tier-1 provider}.

To account for the difference in link bandwidth that likely exists between links serving larger ASes compared to smaller ones, we also assigned weights based on CAIDA's AS rank~\cite{caida_asrank}. This rank orders ASes by their customer cone size. An AS's customer cone is the set of ASes that are reachable by customer links from the AS~\cite{luckie2013relationships}. While CAIDA's AS rank is in descending order (rank 1 having largest customer cone) we invert the order for weighting purposes so that higher link weights indicate larger endpoint AS customer cones. To capture link capacity as a function of AS endpoint customer cone size, we use both the average and maximum rank (of link AS endpoints) as edge weights. The results demonstrate that within the set of graphs with the same unweighted minimum cut there exists widely different difficulties for attackers attempting to disconnect an AS. In fact, a large plurality of steered/poisoning AS pairs require a cut equivalent to one link between ASes with an average AS rank double that of the average AS rank (or two links between ASes of average rank). A majority require a cut at least twice as large, implying that \textit{bottlenecks reside on edges touching large ASes}.

\subsubsection{Return Path Diversity and Security Impact}

For Nyx, our findings agree that \action can reach alternative paths. While our betweenness results show the same ASes appear often on multiple steered paths, our reproduction of Nyx shows that in more than 60\% of cases there exists at least one steered path that avoids an arbitrary AS from the original return path. Therefore, Nyx may help the poisoning AS when it is an impacted bystander or when the adversary is targeting the Internet as a whole. Our min. cut measurements reveal that bottlenecks occur in these steered paths, but it is unlikely they are on the weakest links. This means that an adversary strategically targeting the poisoning AS could target the min. cuts, but must work harder to disconnect a Nyx AS over others. In a similar manner, operators can leverage our insights to gain insight into the types of available paths to use after a set of real-world link failures.

For censorship tools such as Waterfall~\cite{Nasr:2017fy}, the success shown for \action presents issues. These systems now must consider attacks similar to Schuchard et al.'s RAD attack~\cite{RAD}. However, our centrality results reveal a significant betweenness, demonstrating that while alternative return paths exist, on average these paths transit a particular set of ASes that can not be steered around. The min. cut results further buttress this result, and indicate strategic locations where censorship circumventors could place decoy routes to prevent a routing-capable adversary from routing around them with poisoning. Some work already approaches finding more diverse paths~\cite{Nasr:2016tp,Bocovich:2016vu,Gosain:2017ty}, but these systems also do not consider adversaries which can steer traffic around decoys on the return path. We suggest future studies examine poisoning from routers in censoring nations (e.g. China or Iran).

\begin{figure*}[ht!]
    \centering
    \subfloat[ROC curves for different models predicting ASes that can execute return path steering]{\label{fig:roc-curves}\includegraphics[width=0.31\textwidth]{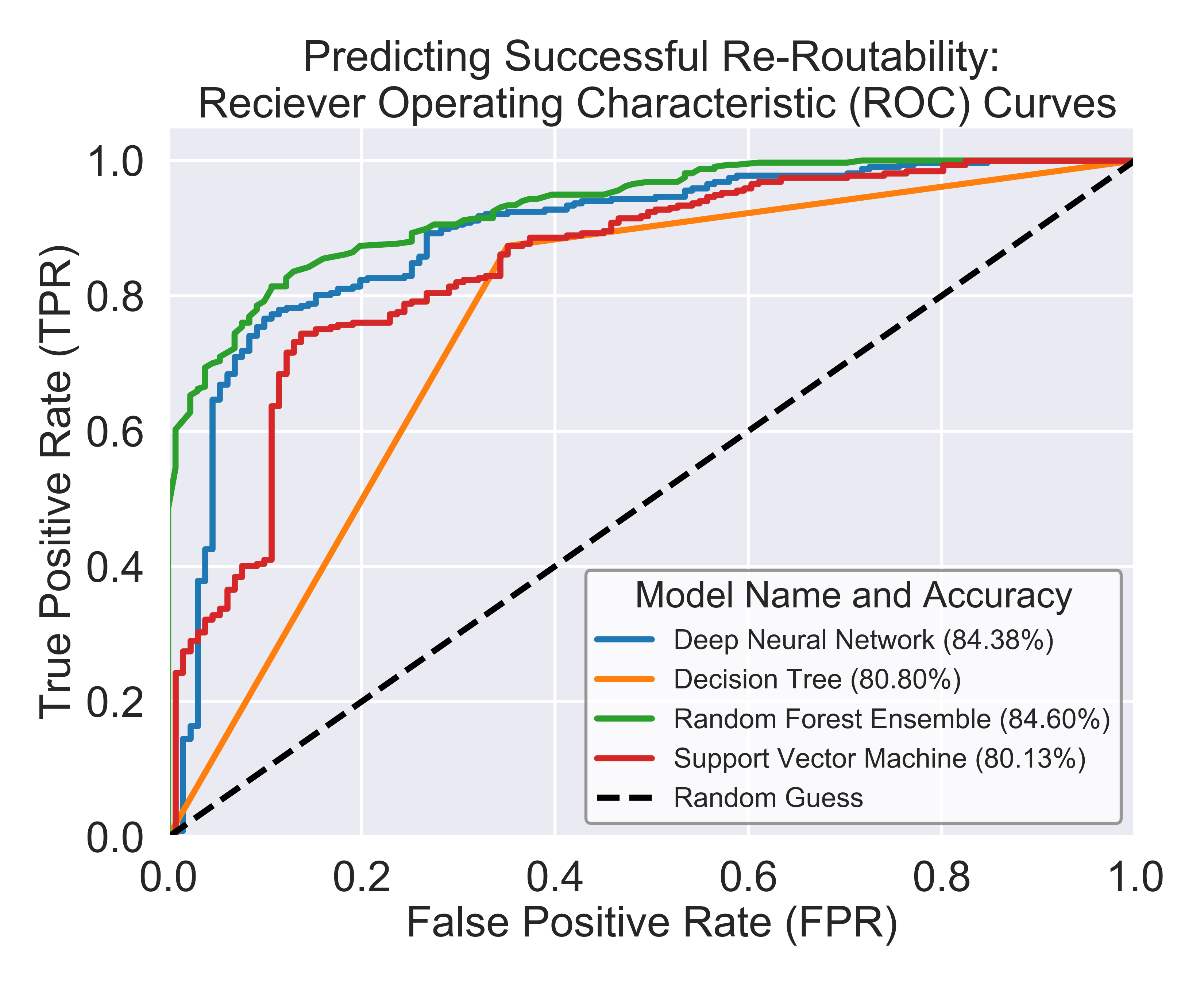}}
    \hspace{0.20cm}
    \subfloat[Features analyzed with Principal Component Analysis, where Index 2 is the Poisoning AS's Next-Hop AS Rank]{\label{fig:feature-importance}\includegraphics[width=0.31\textwidth]{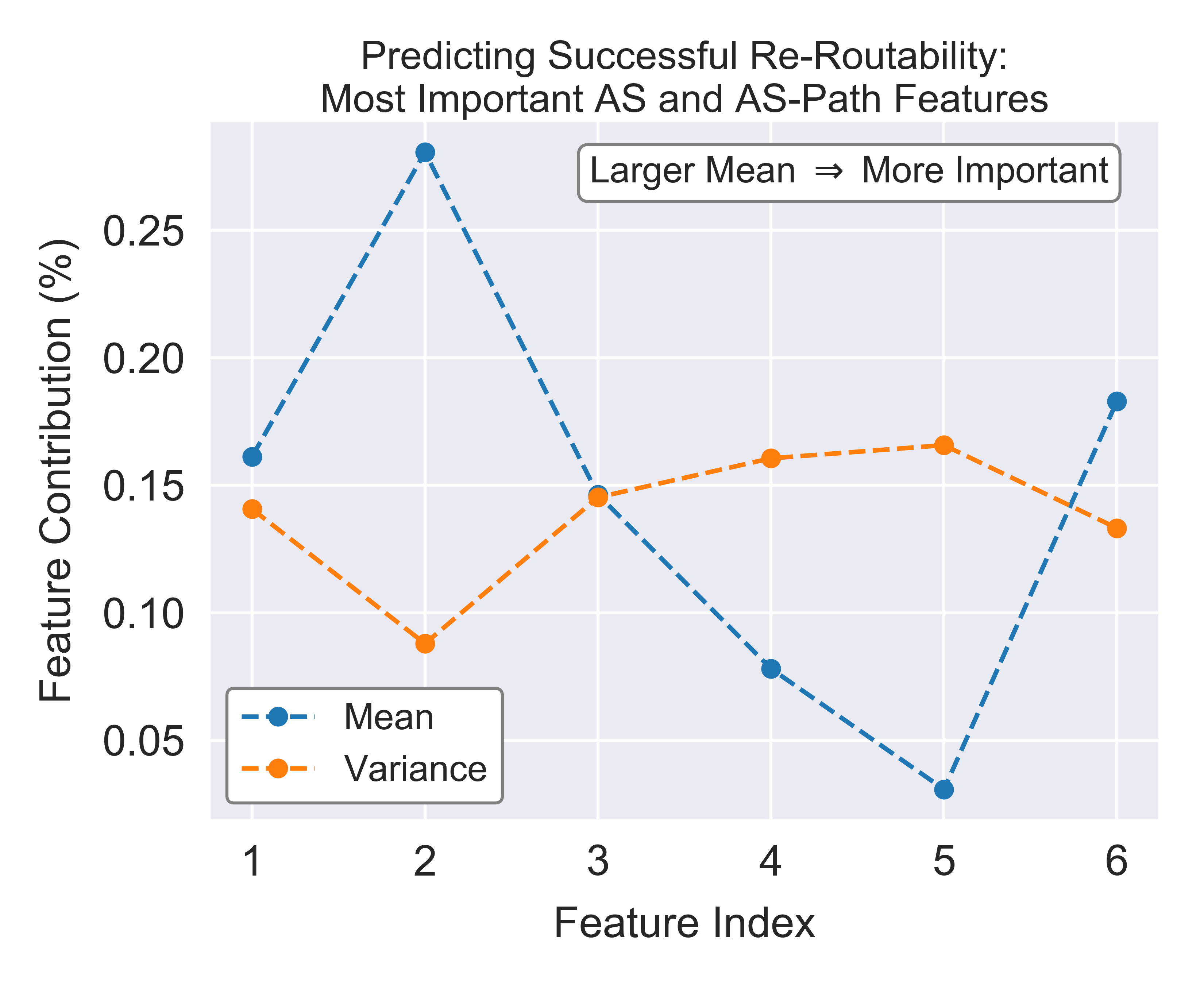}}
    \hspace{0.20cm}
    \subfloat[Distribution of the Poisoning AS Next-Hop AS Rank vs the outcome of the return path steering]{\label{fig:feat-dist}\includegraphics[width=0.31\textwidth]{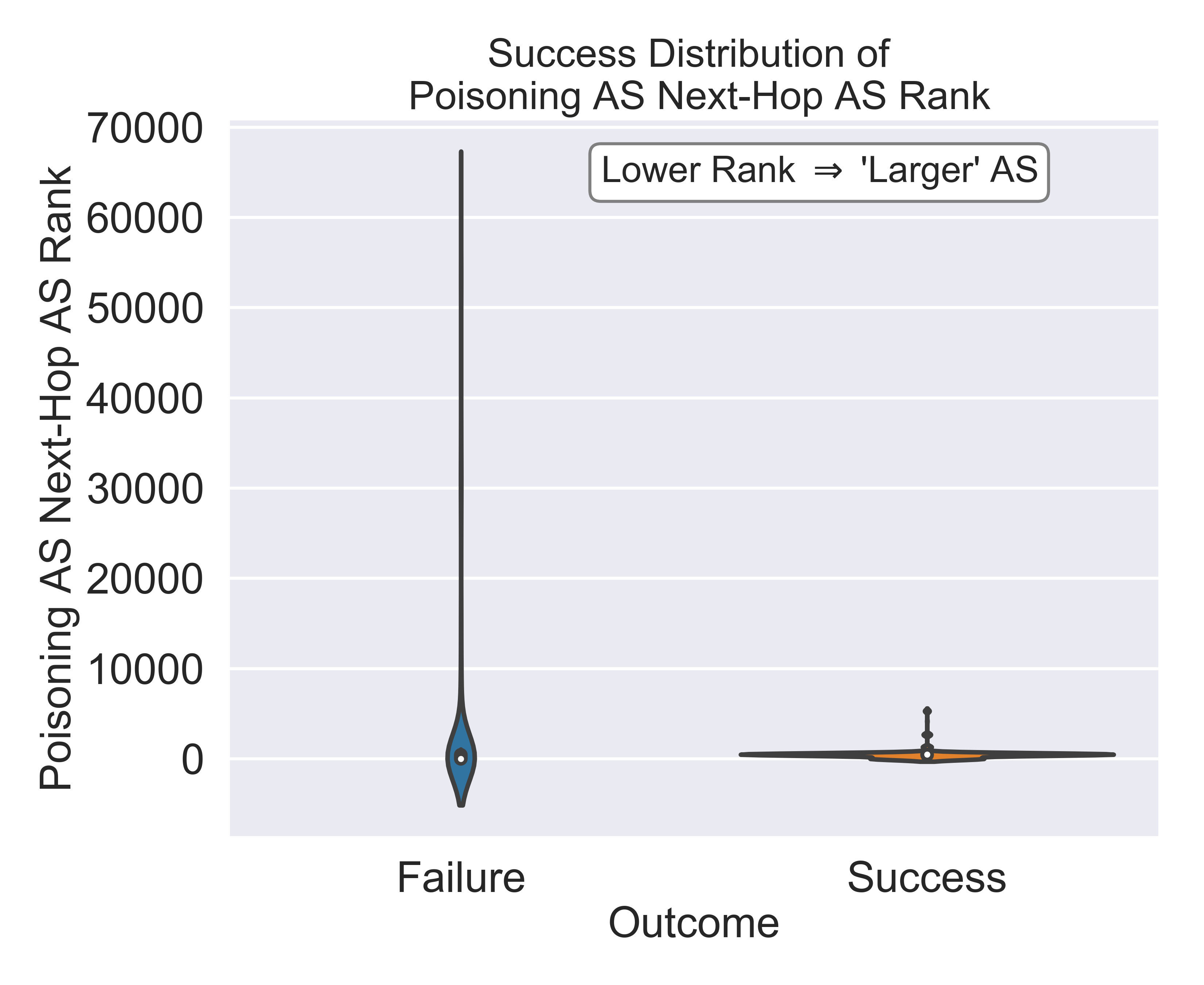}}
    \caption{Predicting successful return path steering with both public and experimentally-derived path-based features: 1) Distance on original path from poisoning AS to steered AS, 2) poisoning AS's next-hop AS Rank, 3) steered next-hop AS rank, 4) original path average edge betweenness, 5) steered AS Rank, and 6) original path average latency (over all hops)}
    \label{fig:mlgraphs}
    \vspace{-20pt}
\end{figure*}

\subsection{Predicting Successful Steering}{\label{predicting-ml}}

To understand which ASes can execute \action most successfully, we constructed a set of statistical models. These models 1) predict which ASes can successfully steer traffic with poisoning and 2) determine the most important predictors for success of \action. Using the entire 1,460 sample dataset, we extract the following features from the real world active measurement data: distance on original path from poisoning AS to steered AS, poisoning AS's next-hop AS Rank, the steered next-hop AS rank, original path average edge betweenness, steered AS Rank, and original path average latency (over all hops). We selected these features based on properties that can be easily determined using standard traceroutes and by referencing open datasets such as CAIDA's AS Rank~\cite{caida_asrank}.

We first split the data into a 70/30 train-test-split. Then we scale the data by removing the mean and scaling to unit variance. In total, we employ 4 models: 4-layer fully-connected neural network, decision tree classifier, random forest ensemble classifier, and support vector machine. After fitting the data, we test the models with a 10-fold cross-validation. Then, we plot Receiver Operating Characteristic curves in Figure~\ref{fig:roc-curves}, which show the success of a given AS at \action. Specifically, the curves show the true positive rate vs. false positive rate distribution across models.

Overall, the models perform strongly. At 80.80\% accuracy, the decision tree classifier both trains and tests new samples the quickest at < 1 second \emph{and} is the most explainable. Explainability of machine learning models is critical here, since operators must inform their network administrators \textit{why or why not} their network is fit for employing \action. Using only the feature vectors and their distribution, we now examine the features that express the most variance.

Figure~\ref{fig:feature-importance} shows a Principal Component Analysis (PCA) algorithm used to rank all features by their mean and variance. The features with higher means indicate more important properties of the poisoning AS, steered AS, and pre-steering AS path. We find the most important predictor is the \textit{next-hop AS rank of the poisoning AS}. As the number of available links to steer onto increases, the poisoning AS finds more unique paths. We can see this by examining Figure~\ref{fig:feature-importance} Feature Index 2. The successful cases evolve from influential ASes as the poisoning ASes' next-hop provider or peer. By drilling down further into the distribution, we see in Figure~\ref{fig:feat-dist} unsuccessful cases clustered around much smaller ASes. Note that path lengths average around 4 hops on the current Internet. In cases where a poisoning AS can not steer through the available paths at its next-hop, other diverse AS choices should exist at the later hops. Perhaps counter-intuitively, the least important predictor is the AS rank of the steered AS. This indicates that the relative influence, or size, of the steered AS \emph{does not} affect a poisoning ASes' ability to steer them.

\subsection{Security Ramifications and Takeaways}

Our findings on the feasibility of steering return paths impact all security systems mentioned in Section~\ref{impact}, including Nyx, LIFEGUARD, RAD, Waterfall of Liberty, and Feasible Nyx~\cite{RAC, KatzBassett:2012gd, RAD,Nasr:2017fy,tran2019feasibility}. Notably, the claims made by systems that leverage BGP poisoning are more in line, but not an exact match, with the behavior of the live Internet. Nyx can successfully execute its re-routing defense using poisoning, though with 30\% less success than simulations show. In particular, poisoning-enabled victim ASes can defeat link-flooding adversaries that target the victim's provider links by executing Nyx to re-route onto alternate, uncongested ASes. Besides demonstrating that BGP poisoning does function in practice for many cases, these experiments also help underscore the need for real-world experiments when validating system design.

From the perspective of censorship, the feasibility of BGP poisoning allows censors to leverage RAD~\cite{RAD} to thwart the efforts of those wishing to avoid censorship with decoys or advanced defenses like Waterfall~\cite{Nasr:2017fy}. However, as we saw with Nyx, we do see that BGP poisoning functions less often than simulations would lead us to believe in specific cases.  In the next section, we explore policies such as AS-level filtering which hamper the effectiveness of BGP poisoning, yet open the door for systems such as Waterfall to function effectively.
\section{Extent and Impact of Filtering}{\label{filtering}}

In this section we present experiments that uncover ASes throughout the Internet which refuse to propagate BGP paths with poisoned ASes prepended. We term this effect \textit{poison filtering}. We present evidence for how often ASes conduct poison filtering, a behavior that impacts the success of BGP poisoning. We explore how many ASes propagate poisoned routes, how long of poisoned paths can be propagated, and additionally conduct a rigorous graph-theoretical analysis of the specific ASes by size inferred to be filtering long poisoned paths. We also attempt to reproduce recent work by Tran \& Kang et al.~\cite{tran2019feasibility} who used a dataset gathered through passive measurement (as opposed to active BGP measurements). In this analysis, we yet again demonstrate that simulation or passive measurement is \emph{not enough} to empirically determine the behavior of the Internet.

\subsection{Filtering of Poisoned Advertisements}{\label{filtering-by-degree}}

Systems which depend on \action need the ability to avoid ASes of a variety of sizes. Since a poison is essentially a lie about the AS level path, it is natural to ask if ASes disregard lies about large ASes. This type of poison filtering would prevent systems using \action from avoiding key ASes in the topology. In order to explore this, we measured the ability to propagate poisoned routes containing various sizes of poisoned ASes.

\subsubsection{Experimental Design}

First, we randomly sampled 5\% of ASes seen in BGP updates from January 2018 by their degree of connectivity. In cases (like Cogent) where an AS has a unique degree of connectivity, we sample just that AS. However, when many ASes share a degree (e.g., 3), we sample 5\% of those ASes uniform at random. With these ASes, we proceeded to advertise poisoned paths with one sampled AS prepended as the poison per advertisement. This announcement would appear as $AS_P, AS_F, AS_P$, where $AS_P$ is our poisoning AS or measurement point, $AS_F$ is the AS being tested for poison filtering. Prior work has found that the relative connectivity of an AS often determines its reaction to anomalous Internet events~\cite{Oliveira:wn,Bush:2009vc} due to larger ASes necessarily enforcing certain policies based on the customers it serves. For each iteration, we initially announce the normal, non-poisoned prefix to all providers and peers connected to the university's AS. After waiting 40 minutes for BGPStream to continuously pull from update collectors in batches, we then fetch all observed updates from the prior 40 minutes, though our updates actually propagated within 30 seconds when observed from the actual collector based on update timestamps. We then measured how many unique ASes were observed advertising the original announcement.~\footnote{BGP convergence happens nearly instantly with poisoned routes, see LIFEGUARD~\cite{KatzBassett:2012gd}.}

With a baseline taken for the non-poisoned announcement per prefix, we repeated this process, but now poisoning ASes by degree from high-low in the path example shown earlier. Again, we wait 40 minutes before collecting all updates and additionally \textit{implicitly withdraw} each poisoned path after each iteration. We then compute the normalized percent of ASes propagating the poisoned paths. This measures the fraction of ASes advertising specific poisoned paths versus those who advertised the non-poisoned baseline path from the equivalent poisoning AS and poisoned prefix. If an AS propagated the non-poisoned path from our AS, and they also propagated a poisoned path, then the normalized percent is higher. In other words, this metric illustrates the percentage of ASes from our random sample that \textbf{do not} employ poison filtering.

\begin{figure}[!ht]
	\centering
\includegraphics[width=0.90\columnwidth]{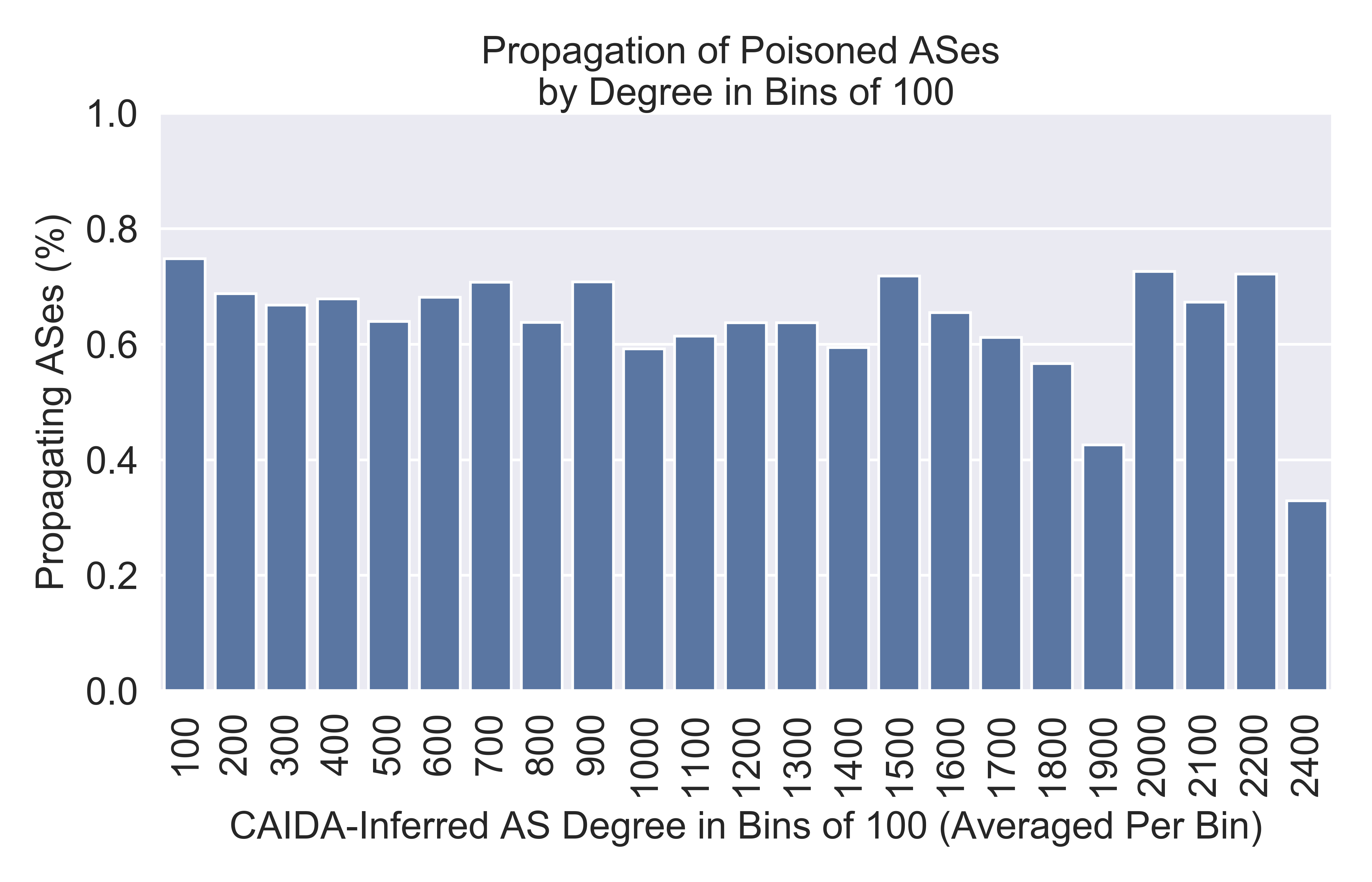}
	\vspace{-8pt}
	\caption{Filtering of AS paths increases as the poisoned AS increases in degree, an approximation for its influence on the Internet}
	{\label{fig:bydegree-fig}}
	{\vspace{-20pt}}
\end{figure}

\subsubsection{Results and Discussion}

The results of this measurement are shown in Figure~\ref{fig:bydegree-fig}. We have aggregated the normalized propagation percentages by AS degree into averages in bins of AS degrees from 0 to 99, 100 - 199, ..., 2300 - 2399. We observe that for AS degrees of less than 2,500, the ASes accepting and propagating the poisoned ASes is roughly the same, with between 70\% to 80\% of ASes continuing to propagate poisons. We did not show the most connected ASes in Figure~\ref{fig:bydegree-fig} due to their outlier status; instead, the top 10 ASes by degree are shown with their propagation data and other relevant AS metadata in Table~\ref{table:top-degree-filters}. Notably, the largest degree AS is Hurricane Electric, a nearly Tier-1 AS. at 7,064 degree. Hurricane Electric has roughly 20\% propagation compared to ASes with under 2,500 customers at roughly 70\% propagation. In fact, the extent to which ASes refuse to propagate high degree poisons is confined to a very small sample of high-degree ASes. Only 4 have a propagation percentage of less than 30\%, with AS degrees of 2,538, 4,980, 5,352, and 7,064.

\begin{table*}[!ht]
\centering
\caption{Top 10 ASes by Degree and their normalized percent of ASes propagating paths with these ASes poisoned}
\resizebox{0.90\textwidth}{!}{%
\begin{tabular}{cccccc}
\toprule
\textbf{Rank by Degree} & \textbf{ASN and Name}     & \textbf{Degree} & \textbf{Number of Customers} & \textbf{Registered Country by ASN} & \textbf{Normalized Propagation Percentage} \\ \midrule
1                       & 6939 - Hurricane Electric & 7064            & 1202                         & United States                      & 11.9\%                                     \\
\rowcolor[HTML]{EFEFEF} 
2                       & 174 - Cogent              & 5352            & 5272                         & United States                      & 11.6\%                                     \\
3                       & 3356 - Level 3            & 4980            & 4898                         & United States                      & 11.6\%                                     \\
\rowcolor[HTML]{EFEFEF} 
4                       & 24482 - SG.GS             & 3382            & 24                           & Singapore                          & 96.1\%                                     \\
5                       & 3549 - Level 3 GBLX       & 2538            & 2446                         & Unites States                      & 11.6\%                                     \\
\rowcolor[HTML]{EFEFEF} 
6                       & 7018 - AT\&T              & 2373            & 2330                         & United States                      & 0.05\%                                     \\
7                       & 58511 - Anycast           & 2351            & 13                           & Australia                          & 60.1\%                                     \\
\rowcolor[HTML]{EFEFEF} 
8                       & 49605 - IVO               & 2193            & 11                           & Italy                              & 66.7\%                                     \\
9                       & 8492 - OBIT Ltd.          & 2153            & 46                           & Russia                             & 71.4\%                                     \\
\rowcolor[HTML]{EFEFEF} 
10                      & 8220 - COLT Tech. Grp.    & 2143            & 716                          & United Kingdom                     & 78.2\%                                     \\ \bottomrule
\end{tabular}%
}
\label{table:top-degree-filters}
\vspace{-10pt}
\end{table*}


First, systems such as Nyx~\cite{RAC} and RAD~\cite{RAD} assume all ASes \emph{do not} conduct poison filtering. We present evidence that significant parts of the Internet do not allow poisoned routes to propagate, especially for the small amount of ASes with degrees greater than 1,000. This finding exemplifies the reason why systems such as Nyx do not find the nearly limitless available paths in practice as what is shown via CAIDA data. To that end, future systems employing BGP poisoning for defensive or offensive purposes should not assume all available paths can be steered onto.

For decoy-routing systems, decoy routers should be placed on AS paths that because of filtering, the adversary can not easily steer said path around the decoy. In scenarios where decoy placement leverages these strategies, the censors may face a losing scenario. Also shown in Table~\ref{table:top-degree-filters}, the number of customers an AS has seems to indicate the extent of poison filtering. For example, despite AS 24482 (an ISP in Singapore) having the 4th highest level of AS connectivity, it only provides direct customer transit to 24 ASes. Accordingly, this Singaporean AS has a much higher propagation percentage relative to ASes with similar degree but more customers. In the case of AS 24482, the non-transit ASes pumping up the AS degree may be peers. Clearly, while paths with larger ASes seen in poisoned paths may be filtered more often, it is not always the case based on AS 24482. With over 3000 ASes reported as connected by CAIDA~\cite{CAIDA}, the amount of propagation was still 96\% of a normal non-poisoned path.

\subsection{Filtering of Long Poisoned Paths}{\label{long-paths}}

Our next experiment investigates the maximum amount of poisoned ASes a poisoning AS can spread throughout the Internet via successively longer path lengths. In existing security systems, Nyx~\cite{RAC} advertises long poisoned paths to avoid dragging along non-critical traffic when steering remote ASes around congestion. RAD~\cite{RAD} and censorship tools using BGP poisoning must rely on many poisons to steer traffic around decoy routers. AS relationship and policy inference methods could use our path steering algorithm from Section~\ref{path-enumeration} coupled with longer poisoned paths to explore broader AS-to-AS business relationships~\cite{Anwar:2015tk}. Congestion discovery systems could also benefit from greater topological visibility.

To that end, we have conducted what we believe is the \textit{most exhaustive measurement of maximum path length on the Internet}. This experiment provides valuable information on whether common models of routing hold in practice. Though the BGP specification~\cite{RFC4271} does not place an upper bound on path length, the BGP best practices RFC~\cite{rfc7454} recommends that excessively long paths should be filtered. Furthermore, statistics from the APNIC routing registry~\cite{potaroo} show most maximum path lengths observed well-under what should be possible. Many Cisco forum posts also hint at operators that assume all paths are filtered over 50 in length. Fortunately, we were able to conduct our experiment from the university AS with permission over two large ISP transit links, without the path length restrictions of PEERING. The university AS's providers have the explicit policy of filtering BGP advertisements longer than 255 hops. Therefore, even though paths may extend beyond this in some router's policies, we can only observe the propagation of path lengths up to 255.

\subsubsection{Experimental Design}

Similar to the poison-filtering approach in Section~\ref{filtering}, we first announce a normal baseline path with no poisons. After collecting the baseline number of ASes advertising the normal path and withdrawing the baseline advertisement, we then iteratively poison paths of increasing lengths in intervals of 40 minutes, from 1 poisoned AS prepended to the path to 135 poisons by one at a time. Once we reached 135 poisons, we shifted to poisoning in successive iterations of 5, going from 135 to 500. After every iteration of path length increase, we implicitly withdrew the prior advertisement. During propagation throughout the Internet, we collect all BGP updates from collectors managed by BGPStream~\cite{Orsini:2016ug}, which we again use to measure the normalized percentage of ASes propagating the poisoned paths. In practice, the path would look similar to the path in Equation~\ref{equ:long-path}, where $AS_{I}$, $AS_{J}$, and $AS_{K}$ are normal ASes forwarding the prefix; $AS_{Orig}$ is the poisoning AS; and $AS_{P_{1}}$ through $AS_{P_{n}}$ are the prepended poisons.

\vspace{-18pt}
\begin{equation}\label{equ:long-path}
	\resizebox{0.4 \textwidth}{!}{
		$AS_{I},\ AS_{J},\ AS_{K},\ AS_{Orig},\ AS_{P_{1}},\ AS_{P_{2}},\ AS_{P_{3}},\ ...,\ AS_{P_{n}},\ AS_{Orig}$
	}
\end{equation}
\vspace{-18pt}

We conducted this experiment with two sets of ASes to prepend: 1) randomly sampled, in-use ASes from the CAIDA topology to most closely mirror a poisoned path needed for \action, and 2) using the university AS as a self-prepend. We ensured part of the in-use AS sample included both ASes on the edge of the topology (those with no customers), as well as transit ASes small and large (those with more than 5 customers) according to prior classifications of AS types by UCLA~\cite{Oliveira:wn}.

\begin{figure}[]
    \centering
	\includegraphics[width=0.80\columnwidth]{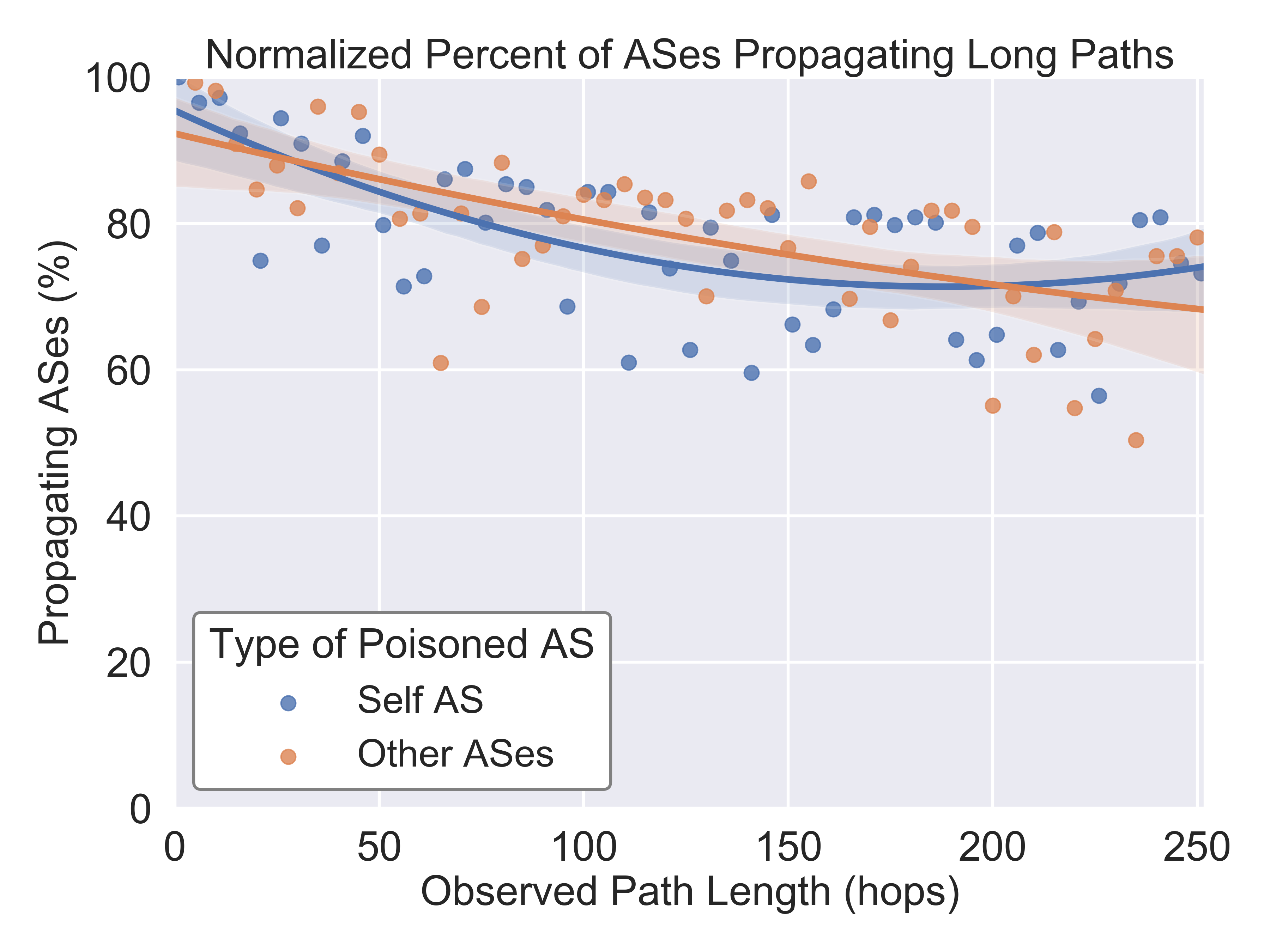}
	{\vspace{-5pt}}
	\caption{With paths up to 250 in length, we found over 80\% of ASes treated 250-length paths the same as normal paths (Regression Fit of Order 2)}
	{\label{fig:longpath-both}}
	{\vspace{-25pt}}
\end{figure}

\subsubsection{Results and Discussion}

Displayed in Figure~\ref{fig:longpath-both} for both the randomly sampled ASes from CAIDA and for the self-prepended university AS, we present a rigorously evaluated upper bound on the max path length of the Internet of \textbf{251}. This path propagated to over 99\% of the Internet when including customer cones of AS's forwarding the path. This included highly connected ASes such as Level 3 and Cogent. Figure~\ref{fig:longpath-both} matches an operator's intuition that as paths grow longer, they are less accepted throughout the Internet, though still roughly 75\% of BGP collectors observed the longest path lengths detected.

With this information, systems such as Nyx~\cite{RAC} now have an upper bound for the amount of poisoned ASes usable for path lining, which was estimated with passive, not active, measurements in Tran \& Kang et al.'s re-routing feasibility study~\cite{tran2019feasibility}. Since Smith et al. did not limit the poisons, our reproduction of Nyx earlier incorporates this poison limit, finding less success overall when steering return paths. When re-routing around localized failures, as Katz-Bassett et al.~\cite{KatzBassett:2012gd} did between Amazon AWS instances in LIFEGUARD, this maximum length limits the amount of path steering in practice that can be achieved. There are implications for RAD~\cite{RAD} and other decoy routing adversaries as well: the more poisons possible, the harder Waterfall~\cite{Nasr:2017fy,Bocovich:2018wu,Nasr:2016tp} must work to place decoys.

\subsection{Which ASes Filter Long Paths}{\label{long-path-filtering}}

\subsubsection{Filtering Inference Algorithm}

Here we investigate which ASes are filtering paths based on data collected in the prior experiment. We develop a new inference algorithm to discover which ASes filter long poisoned paths based on a comparison of paths received by route collectors at each advertisement of successively greater length. First, we build a directed acyclic graph $D$ of all paths $p$ observed on paths from the university AS to collectors. The nodes of $D$ are ASes appearing on paths; edges represent links between them. Next, for each advertisement $i$ of successively greater path length, we build a set of ASes $A_i$ composed of all ASes appearing on our advertised paths that reached route collectors. Finally, we remove all $a\in A_i$ from a copy of $D$, creating $D_i$. For each weakly connected component remaining in $D_i$, we learn that 1) at minimum, the roots of each component filtered the advertisement, and 2) at maximum, all AS nodes $a\in D_i$ filtered it. Using this method, we iteratively build maximum and minimum inferred filtering AS sets for every path length in our experiment.

\begin{figure*}[ht!]
    \centering
    \subfloat[Tier-1 and Large ISPs]{\label{fig:tier1largeisp-filtering}\includegraphics[width=0.31\textwidth]{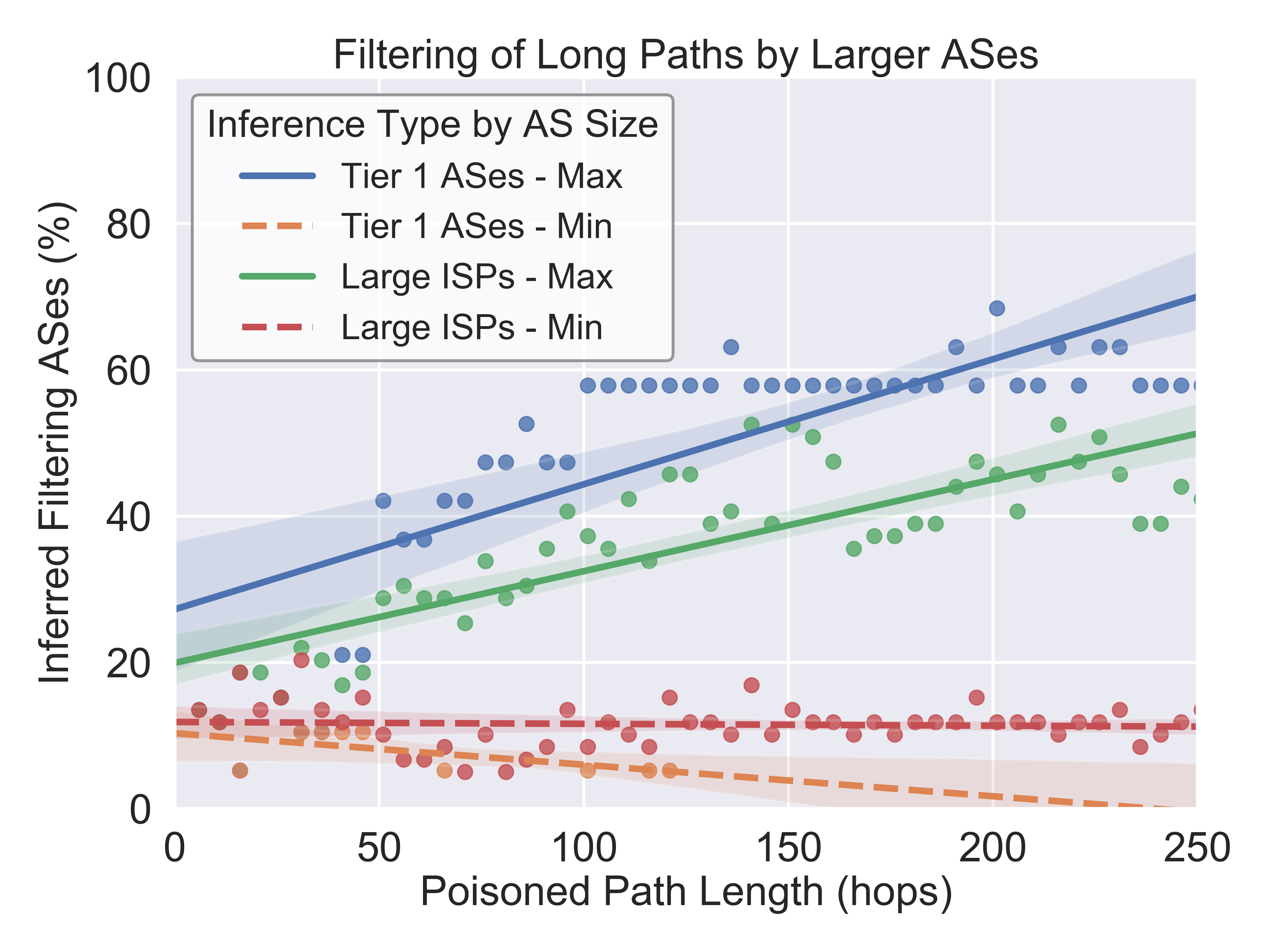}}
    \hspace{0.20cm}
    \subfloat[Small ISPs and Stub ASes]{\label{fig:smallstub-filtering}\includegraphics[width=0.31\textwidth]{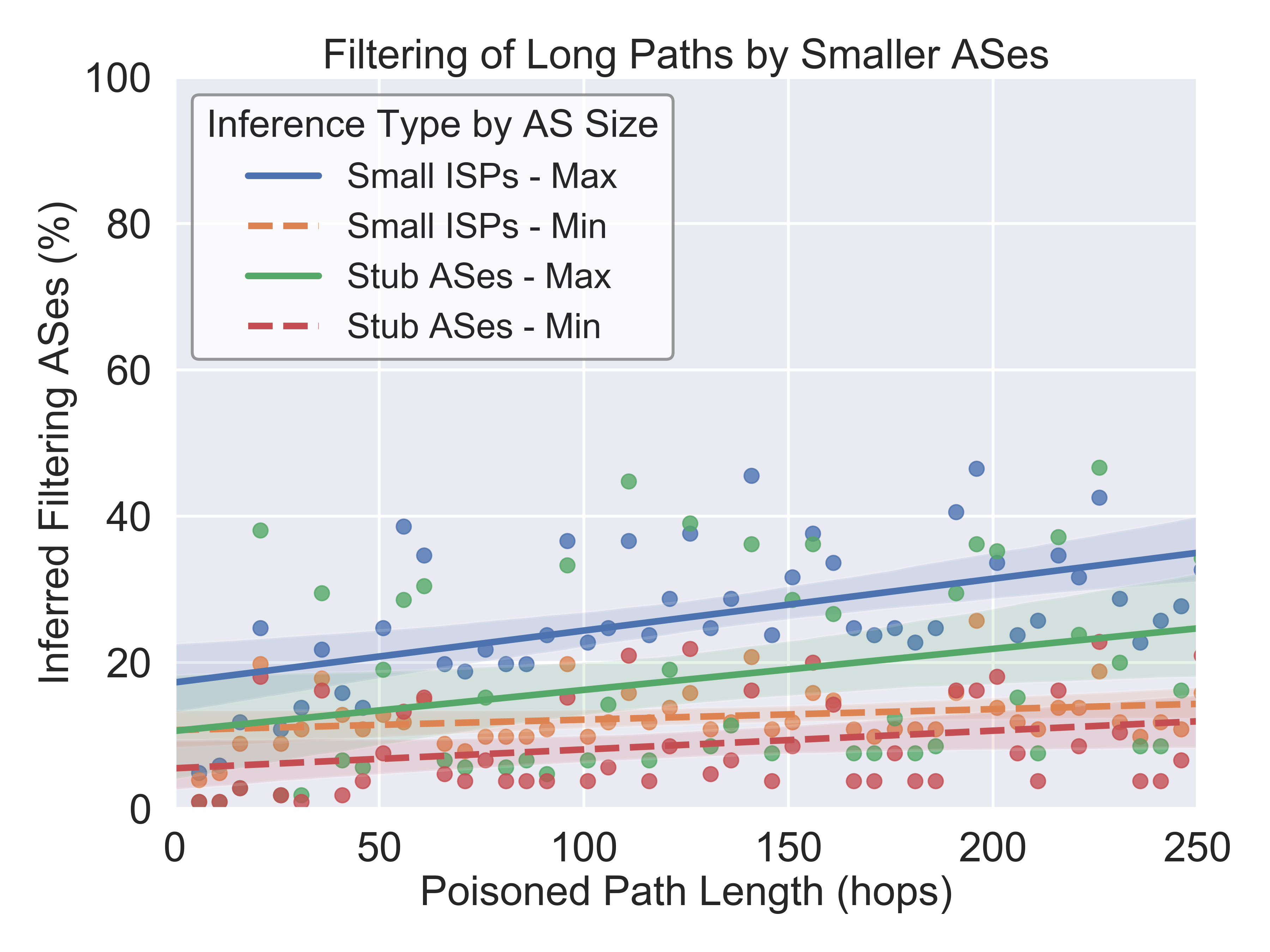}}
     \hspace{0.20cm}
    \subfloat[MANRS vs non-MANRS ASes]{\label{fig:manrs-filtering}\includegraphics[width=0.31\textwidth]{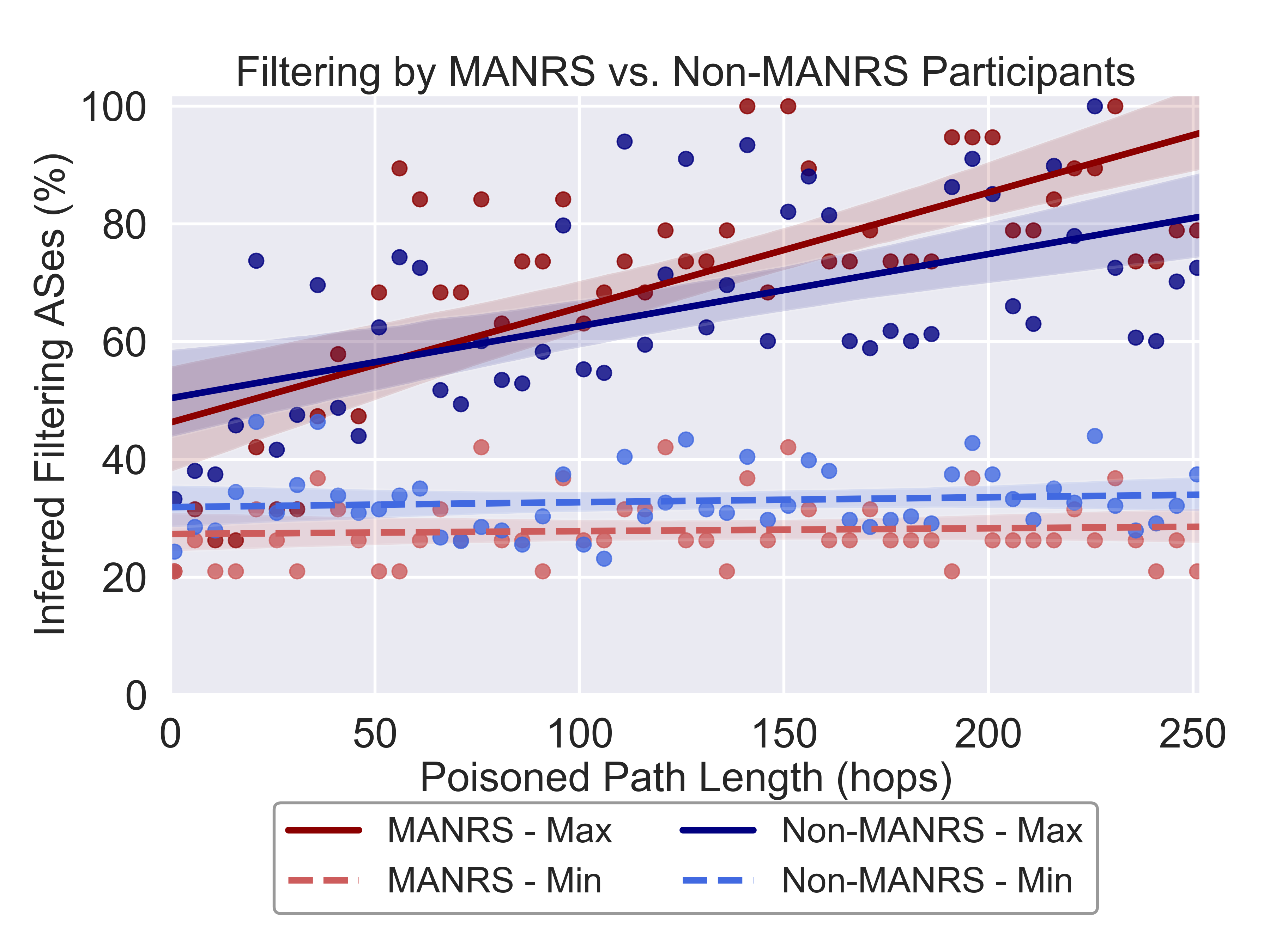}}
    \caption{Minimum and maximum inferred filtering for ASes classified by tier and MANRS membership, each with an regression fit}
    \label{fig:inferred-filtering}
    \vspace{-20pt}
\end{figure*}

\subsubsection{Results and Discussion}

Our results are grouped using the aforementioned, widely-adopted AS classification scheme presented in~\cite{Oliveira:wn}. ASes are divided into Tier-1 (can transit traffic to all ASes without compensation and form a clique), Large ISPs with over $50$ customers, Small ISPs with between $5$ and $50$ customers, and Stub ASes, those with less than $5$ customers. Figure~\ref{fig:tier1largeisp-filtering} displays our results for Tier-1s and Large ISPs; Figure~\ref{fig:smallstub-filtering} gives the same information for Small ISPs and Stub ASes. Naturally, the ephemeral structure of the Internet topology introduces noise into our results. Additionally, it is more difficult to draw conclusions about Tier-1 and Large ISP filtering behavior using our method, as the minimum and maximum inferences diverge significantly. This is likely due to advertisements being filtered before reaching these ASes as they propagate outward from the university AS. So, these ASes are rarely the root of the weakly connected components used to infer minimum filtering, and we conjecture that the true filtering rate for these classes is closer to the maximum inference.

Overall, the results indicate that Tier-1's and Large ISPs filter long paths more aggressively than Small ISPs and Stub ASes, and that AS filtering policies are highly fragmented. In a feasibility study on Nyx/RAC by Tran et al.~\cite{tran2019feasibility}, the authors utilize a distribution of observed path lengths from passive measurement to hypothesize about AS filtering rates. In short, they suggest that some filtering occurs on paths of length $30$ - $75$, no increase in filtering occurs between $75$ and $255$, and paths of length $255$ or greater are almost universally filtered. We were limited by university AS provider policy from experimenting with paths over length $255$, but their findings align well with our own for Small ISPs and Stub ASes. For the larger ASes, our experiments indicate that the rate of filtering does in fact increase after a path length of $75$. Additionally, our results capture the intuition that larger, more influential ASes should filter often. We find that of all tiers of ASes, the Tier 1 ASes filter most, while larger ISPs filter less but close behind Tier 1 ASes shown in Figure~\ref{fig:tier1largeisp-filtering}. Finally, small ISPs and stub ASes filter very little as shown by Figure~\ref{fig:smallstub-filtering}.

\subsection{Case Study: Filtering by an ISP-driven Working Group}

MANRS~\cite{MANRS}, Mutually Agreed Norms for Routing Security, is a global Internet routing security initiative that develops and publishes best practices for network operators. Path filtering is one area of concern for MANRS, and they publish standards for following RPKI and other BGP security mechanisms that member ASes are expected to implement. The 120+ MANRS ASes represent a distinct set of ASes that intuitively should be most likely to filter BGP advertisements similar to poisoned updates. They include Cogent, Charter Communications, CenturyLink, and Google.

In Figure~\ref{fig:manrs-filtering}, we display the results of the same filtering inference algorithm used in the previous section, with results divided by MANRS and non-MANRS ASes. We observe a significant deviation in the inferred filtering range between MANRS and non-MANRS ASes, suggesting that MANRS operators may implement tighter filtering policies. This key result indicates that an ISP's participation in an Internet consortium such as MANRS may actually correspond with stricter implementations by the operators responsible for day-to-day network activity and filtering policy, rather than aligning with MANRS only at an organizational level.

\subsection{Security Ramifications and Takeaways}

Our findings on poison filtering impact all security systems mentioned in Section~\ref{impact}, including Nyx, LIFEGUARD, RAD, Waterfall of Liberty, and Feasible Nyx~\cite{RAC, KatzBassett:2012gd, RAD,Nasr:2017fy,tran2019feasibility}. While the results of Section~\ref{path-enumeration} show that BGP poisoning \emph{broadly} functions, these experiments reveal that AS-level filtering in portions of the Internet results in \emph{specific topological locations} where poisoning \emph{does not} function. As a result, systems like Waterfall, which depend on BGP poisoning not to function should seek out those locations for deployment. By seeking out topological regions of the Internet where poisoning is not effective, some of which were described earlier, censorship circumvention systems can avoid RAD adversaries wishing to route around decoys.

From the perspective of DDoS, Nyx cannot easily route around DDoS in the filtered regions of the Internet. This filtering contributes to the weakened effectiveness of Nyx in practice that we saw in the prior section.  The filtering of long paths also affects Nyx by limiting its ability to poison the neighbors of the alternate paths, which prevents dragging along unintended traffic, thus hampering the relief of attack congestion. Notably, these experiments also put a hard real-world limit on Internet path length. At 255 ASes, Internet paths fail to propagate; however, up to 255, paths propagate to 99\% of ASes, unlike what was found with passive measurement in prior studies~\cite{tran2019feasibility}. ASes seeking to propagate long paths should limit the number of poisons used to ~245, which accounts for up to 10 ASes for the actual path. Given that the average path length is between 3 and 4 ASes~\cite{potaroo}, this amount of room is suitable for real-world deployment of poisoning.
\section{Reassessing Reachability}{\label{reachability}}

%

As part of our study, we setup our infrastructure to attempt to reproduce Internet measurements from nearly a decade ago by Bush et al. in Internet Optometry at IMC 2009~\cite{Bush:2009vc}. These measurements from 2009 have a distinct impact on the feasibility of BGP poisoning for security. This section presents our findings for an Internet in 2018 with over 60,000 ASes and estimated 3.8 billion unique users, compared to nearly 25,000 ASes and 1.7 billion users in 2009~\cite{internetusers,potaroo}. Once again, we found cases in this evaluation where common assumptions from the operator community \textit{did not coincide} with actively measured Internet behavior.

\subsection{Declining Presence of Default Routes}{\label{defaults}}

\begin{table}
\centering
\caption{Default Route Findings}
\resizebox{0.90\columnwidth}{!}{%
\begin{tabular}{L{4.0cm}L{4.0cm}}
\toprule
\textbf{Measurement}                                             & \textbf{Number of Instances}             \\ \midrule
Fraction of Total Samples with Only 1 Provider (not multi-homed) & 28.7\% (419 / 1,460 total samples)       \\
\rowcolor[HTML]{EFEFEF} 
Fraction of Total Multi-Homed Samples with Default Routes        & 48.6\% (506 / 1,041 multi-homed samples) \\
Fraction of Transit ASes with Default Routes                     & 26.8\% (196 / 731 total Transit ASes)    \\
\rowcolor[HTML]{EFEFEF} 
Fraction of Stub/Edge/Fringe ASes with Default Routes            & 36.7\% (310 / 845 total Fringe ASes)     \\ \bottomrule
\end{tabular}%
}
\label{table:dr-metrics}
\vspace{-10pt}
\end{table}


\begin{table}
\centering
\caption{/25 Reachability Findings}
\resizebox{0.98\columnwidth}{!}{%
\begin{tabular}{L{0.8cm}L{2.0cm}L{2.3cm}L{2.8cm}}
\toprule
\textbf{Prefix Length} & \textbf{Measurement}    & \textbf{Findings}                  & \textbf{Timespan of Measurement}                 \\ \midrule
/25                    & BGP Observability       & Seen at 21/37 (56.7\%) collectors  & 96 hours of collection                           \\
\rowcolor[HTML]{EFEFEF} 
/25                    & Traceroute Reachability & 31\% reached /25 prefix on average & 7 hours; 5,000 distinct traceroutes every 1 hour \\
/24                    & BGP Observability       & Seen at 34/37 (91.8\%) collectors  & 96 hours of collection                           \\ \bottomrule
\end{tabular}%
}
\label{table:connectivity-results}
\vspace{-10pt}
\end{table}

\subsubsection{Experimental Design}

Default routes exist when an AS has two or more providers and refuses to choose a second provider when the first provider is "removed" from the topology via BGP poisoning. A poisoning AS can theoretically remove the first provider from the steered AS's topology by causing the first provider to drop (and not propagate) its route to the poisoning AS. However, when measured in 2009~\cite{Bush:2009vc}, this did not always occur when poisoning. Default routes were evaluated as part of the earlier return path steering experiments.

\subsubsection{Results and Discussion}
We found 330/1,460 successful poisoning cases exhibited default routes at the steering AS's next hop. Thus, steering traffic onto a second, third, or other provider bordering the remote AS was impossible for these ASes. As shown by Table~\ref{table:dr-metrics}, we examined the properties of these ASes by their transit or fringe status. Default routes existed for 26.8\% of the transit ASes we sampled, those being ASes with 5 or more customers. We also found that 36.7\% of the fringe ASes, or those with less than 5 customers, had default routes. In 2009, 77\% of stub ASes had default routes (out of 24,224 ASes measured from the poisoner with the ping utility). In 2018, we found 36.7\% of stubs had default routes (out of 845 ASes measured from the steered AS with RIPE Atlas probes using traceroute).

Based on the prevalence of default routes, decoy routing systems~\cite{Nasr:2017fy,Nasr:2016tp,Bocovich:2018wu} could optimize placement of decoy systems on the immediate next-hop of remote ASes which a RAD~\cite{RAD} adversary wishes to steer. This approach may yield stronger security guarantees against poisoning-equipped adversaries when the middle of the path exhibits strong compatibility with poisoning. For systems such as Nyx~\cite{RAC}, a steered AS which has a default route may not be able to provide QoS to its customers when under a direct DDoS due to the inability to steer return paths around its next-hop AS that will always be used for return traffic.

\subsection{Growth of /25 Reachability}{\label{connectivity}}

When executing \action, an originating AS utilizing poisoning on a set of their own prefixes can cause the poisoned ASes and their traffic to \textit{lose} connectivity to the poisoning AS's prefixes, because without a less specific covering prefix, the poisoned AS will have no path to the poisoning AS. Fortunately, a poisoning AS can still maintain reachability given a sufficient allocation of prefixes. However, not all ASes benefit from an ample supply of controlled prefixes. To maintain reachability while poisoning, an AS must have both a more specific prefix and a less specific prefix. Current best practice documents recommend ASes filter advertisements longer than /24~\cite{rfc7454}. Therefore, any AS which only has /24 prefixes available will lose reachability to poisoned ASes \textit{unless} it can advertise a /25 as the poisoned prefix and use the /24 as the covering prefix. We searched the BGP RIB for an AS's shortest advertised prefix length and found that 48\% of them only have /24 prefixes advertised, except for ASes that may have had /25's already in the default free zone. For an in-depth examination of BGP prefix delegation and who gets the privilege of many prefixes, see recent work by Krenc et al.~\cite{Krenc:2016wr}. With this necessary primitive for poisoning in mind, we set out to examine the amount of reachability a /24-only poisoning AS could retain to the poisoned ASes.

\subsubsection{Experimental Design}
For the control-plane measurement, we started by announcing a unique /24 across all 8 prefixes. Over the following 96 hours, we collected aggregated BGP updates from BGPStream~\cite{Orsini:2016ug}. This provides a baseline number of ASes propagating the path. We then withdrew each /24 prefix. Next, we announced a /25 prefix from the same set of locations across PEERING and the university AS, collecting the number of updates in the same manner again over 96 hours. With these two sets of ASes, we compute the normalized percent of ASes propagating the /24 versus the /25.

Next we measure data-plane reachability. We announced a /25 from our 8 prefixes. Then we scheduled 5,000 traceroutes, randomly sampled from all Atlas probes, directed to the advertising /25 prefixes every hour for 7 hours. We recorded the number of traceroutes that reached the /25, noting this as the approximate reachability of the /25. We opted not to do the traceroutes for the /24 both due to PEERING being used for other experiments and because we expect a /24 to be reachable except in the case of faulty Atlas probes.

\subsubsection{Results and Discussion}
Our results from these experiments are shown in Table~\ref{table:connectivity-results}. We found that the current data-plane reachability of a /25 is roughly 30\%, while the number of ASes propagating a /25 BGP announcement is over 50\%, or \textit{50x higher than 2009 results} in Bush et al.~\cite{Bush:2009vc}. Notably, our analysis shows that the 48\% of ASes without sufficient \textit{recommended} prefixes to steer traffic actually \textit{can maintain reachability}. We note that our results may not be able to be directly compared with Bush's work due to the use of traceroutes here over ping, but the comparison at least serves as a measuring stick for the growing Internet.

Our findings also have crucial implications for both existing and future security systems. Censors that do not have a less specific prefix than a /24 will be unreachable by ASes affected by path steering for a non-negligible amount of steered ASes. This may dampen a RAD~\cite{RAD} adversary's success in terms of economic means by lost traffic, while strengthening the case for circumvention tools~\cite{Nasr:2017fy,Bocovich:2018wu,Nasr:2016tp}. Though loss of reachability may not be an issue for censorship entities in general, smaller censoring nations may sustain significant economic costs. Smaller nations may have few egress BGP paths to the broader Internet. Any additional sacrificed reachability may have substantial impacts on the businesses and users behind the border. However, the cases where BGP routers propagate a /25 may be enough in some attack scenarios for a RAD adversary, though this is topology-dependent given a poisoning AS. We recommend that future iterations of the decoy routing attack and defense schemes factor in our findings by evaluating their success for classes of ASes with only a /24 prefix. As a defensive mechanism, operators of Nyx or LIFEGUARD~\cite{RAC} must be willing to sustain losses of reachability. Defensive measures must account for these new AS reachability metrics for ASes with few prefixes.

\subsection{Security Ramifications and Takeaways}

Our findings on reachability and default routes impact all security systems mentioned in Section~\ref{impact}, including Nyx, LIFEGUARD, RAD, Waterfall of Liberty, and Feasible Nyx~\cite{RAC, KatzBassett:2012gd, RAD,Nasr:2017fy,tran2019feasibility}. Like filtering, default routes and losses of reachability limit what can be claimed by poisoning-enabled systems such as Nyx, RAD, and LIFEGUARD. From the perspective of DDoS, Nyx functions even with default routes, since attacks are often more than several hops away from the target in the case of Link Flooding Attacks (LFAs)~\cite{RAC}. Furthermore, our finding that 48\% of ASes with only a /24 can in fact use a /25 is significant for Nyx and LIFEGUARD deployers with few IP prefixes. Thus, smaller ISPs can still execute BGP poisoning feasibly and maintain their connectivity via a less and more specific prefix.

 However, our findings again limit these systems in specific cases. For example, decoy routers benefit from RAD~\cite{RAD} being unable to steer return paths due to default routes near the poisoning AS, or AS deploying RAD. Poisoning also impacts the reachability of censoring ASes' prefixes when a less specific prefix is not available. This presents a difficult decision to censors unwilling to shut off partial Internet access to customer traffic. While censors are inhibited by poisoning in some cases, a censor willing to lose reachability to some parts of the Internet \emph{and} aware of decoys not impacted by default routes can still benefit from BGP poisoning. Finally, our discovery that the Internet has significantly changed from 2009 to 2018 with respect to reachability should come with no surprises. Therefore, systems designed in an era of a vastly different Internet topology may need to be re-evaluated again on the live Internet to see whether their claims are still valid.
\section{Discussion}{\label{discussion}}

\subsection{Reproducibility and Continuous Measurements}{\label{reproducibility}}

All experiments conducted for this paper can be replicated using the same public infrastructure we utilized. Distributed traceroutes can be ran using RIPE Atlas~\cite{RIPEATLAS}, other sources such as NLNOG Ring~\cite{scheitle2017large,holterbach2014suitability} and PlanetLab~\cite{planetlab,spring2006using} can also be used. BGP measurements can be conducted by partnering with the PEERING testbed~\cite{schlinker14peering}, which is available for use by operators and academic researchers via an application process. The experiments in this work were conducted in the first half of 2018, but the measurement infrastructure and framework is open source~\footnote{\url{https://github.com/VolSec/active-bgp-measurement}} and can be deployed to conducted continuous measurements from the same or similar vantage points. We hope to work with PEERING and others at UT to carry these measurements out continuously, providing a live, ever-changing look at BGP poisoning behavior on the Internet.

\subsection{Experimental Limitations}{\label{limitations}}

Our experiments have several limitations. First, some ASes will filter poisoned advertisements, and it can be difficult to fully understand what is driving the behavior without having insider information from the ISP's policies. Second, when we see poisoned advertisements not pass through certain ASes, we are inherently unable to know whether the lack of propagation is due to filtering or if the router is invisible from our perspective on the control plane. We cannot distinguish a case where both our provider and their provider is filtering poisoned routes vs. our provider filtering and then our infrastructure not seeing the new route. This results in our measurements of filtering being an \emph{upper bound} on the amount of filtering occurring. Fortunately, continuous measurements would help address several of these limitations.

From a geographic perspective, we did not attempt to measure the differences in our re-routing feasibility experiments from Section~\ref{path-enumeration} geographically. Recent work has found geolocation of routers and AS paths from public and commercial databases to be unreliable~\cite{gharaibeh2017look}; therefore, we focus on the topological differences in poisoning feasibility. Finally, the instability of RIPE Atlas as our distributed traceroute source can be limiting. Traceroute probes from RIPE Atlas suffer from infrequent instability due to the small profile of the probes and the load on the entire measurement network. To adjust for this behavior, we only use Atlas probes that remain stable for the entire experiment. We define stability here as Atlas probes that continue to respond to API requests and return successful responses when requesting AS-level network paths.

\subsection{Recommendations for Re-Examining Other Security Work}{\label{example-systems}}


There are many examples of other security systems from recent literature, while not focused on poisoning specifically, do make assumptions about BGP behavior and for the same policies as the inferred topology of the Internet to hold in practice. Notably, NetHide, which obfuscates traceroutes across the Internet~\cite{217513}, \emph{tests across less than 1\% of Internet ASes of a total 60,000+}, but instead only \~150. Recent studies of defeating BGP hijacking of Bitcoin, including SABRE by Apostolaki et al.~\cite{apostolaki2017hijacking,apostolaki2018sabre}, only test with the inferred CAIDA topology, which we show in this paper \emph{does not match reality when it comes to poisoning}. Less recent work such as RAPTOR~\cite{190964}, a look at routing attacks on Tor, claims that \/24 prefixes are the only prefixes that cannot be defeated by a Tor adversary using RAPTOR, yet we showed earlier that large swaths of the Internet will respect a \/25, allowing Tor to be de-anonymized in some cases. Finally, a study recently published at IEEE S\&P 2020 by Tran et al.~\cite{tran2020stealthier} focuses on advanced Bitcoin re-routing attacks. However, Tran makes direct claims that actively measuring the BGP policies they require for their attack \emph{is difficult} and do not do so, though we find in this work that it is in fact possible to actively measure BGP policies.

We recommend all of these systems, like our re-evaluation of Nyx, to be reproduced on the live Internet before being deployed to protect actual users. Offensive tools should also be tested ethically before being integrated into the threat models of network practitioners. Much like a modern military would not conduct live fire testing of weapons, nor should academic researchers targeting the live Internet not attempt to ethically measure their tool to show whether the attack would stand up to real-world execution. Some systems from past and recent literature test their hypotheses relying on Internet routing behavior beyond simulation. An example offensive tool which tests using real-world BGP advertisements is SICO, which can launch
interception attacks with BGP communities. SICO leverages PEERING~\cite{birge2019sico}, like this paper. Blink, again by Apostolaki et al., which establishes fast connectivity using the data-plane~\cite{227629}, was tested on the live Internet. SCION~\cite{Perrig:2011tz} and Named Data Networking~\cite{zhang2010named}, both proposed "future Internet architectures", are actively deployed on the live Internet.

\subsection{Related Measurements}{\label{related-work}}

In the background, we covered relevant security literature. Here, we cover related Internet measurement research. LIFEGUARD from Katz-Bassett et al.~\cite{KatzBassett:2010wp,KatzBassett:2012gd} and Anwar et al.'s Interdomain Policy exploration~\cite{Anwar:2015tk} use algorithm similar to our \action methodology. They addressed steering return traffic around link failures between Amazon EC2 servers distributed among data centers. However, our algorithm explores greater depths in its breadth-first search of all possible paths from a single remote AS, rather than aggregating paths available from many poisoning ASes. While not directly related to our steering algorithm, work on BGP communities can influence inbound paths similarly to poisoning. Communities in the wild have been studied by Streibelt et al.~\cite{Streibelt:2018tp}.

\section{Conclusion}{\label{conclusion}}

We have presented measurements demonstrating the limitations of leveraging BGP poisoning in the real-world. Notably, we found 77\% of ASes evaluated could be successfully maneuvered onto new, previously unreachable AS-links at some point in the original path. We found that of all types of AS connectivity, only highly connected ASes were strictly incompatible with BGP poisoning. When poisoning an AS, we demonstrated that for roughly 30\% of the Internet, using non-standard IPv4 prefixes will maintain connectivity when no other routable prefixes are available. Beyond connectivity, we investigated default routing on the Internet and found that for 36\% of ASes with only 2 providers, even in cases where the primary provider is poisoned, the AS being maneuvered will continue to route through the poisoned upstream AS. Finally, we established the first upper bound on the maximum AS-path length routable on the Internet via an exhaustive search, discovering that paths of up to 251 in length are accepted by 99\% of the Internet when considering the customer cones of ASes advertising such paths.

We summarize our key results, major takeaways, and security ramifications in Table~\ref{table:experiment-summary} at the beginning of this paper and provide additional takeaway discussion in the prior section. With these findings, we called out specific assumptions that are now either validated or invalidated for systems such as Nyx, RAD, Waterfall, and others. In particular, the placement of decoy routers can effectively quell the impact of a RAD adversary, while poorly placing decoy routers in the core of the Internet leaves censors with vast territory to intercept traffic. For defense systems such as Nyx, we demonstrated that to some extent, the ability to defend against DDoS under their presented bandwidth models hold; however, the assumptions made about the amount of poisons possible overall and the specific ASes on the Internet able to be poisoned require significant changes to their described algorithms in order to see real-world viability in the range of cases our study shows. Beyond Nyx, RAD, Waterfall, and decoy routing, the study we have presented carries implications for potential alleys to better Internet model-building and other Internet simulation/emulation systems that must model the Internet's de facto routing infrastructure.

Most notably from this study, we strongly recommend all future work that targeting the Internet for its real-world deployment go beyond simulation and passive measurement. Researchers should lean into the behavior of network operators and practitioners, intentionally conducting ethical, well-designed experiments that validate simulations in practice. We have shown that inferred topologies alone nor passive measurements can be the only validation in an era where the true behavior of the inherently human-driven Internet deviates strongly from inference. Like many other fields of science, security researcher's proposed hypotheses for building, defending, and attacking distributed systems need to be conducted in the real-world before the proposed work is disseminated into the literature and public discourse.


\section*{Acknowledgment}
We thank Panos Papadimitratos for his assistance shepherding this paper and our NDSS reviewers for their constructive feedback. We also thank the VolSec lab members and Kaleigh Veca for revisions. This material is based upon work supported by the National Science Foundation under Grant No. 1850379.

\bibliographystyle{IEEEtranS}
\end{document}